\documentclass[a4paper]{article}

\usepackage[utf8]{inputenc}
\usepackage[margin=2cm]{geometry}
\usepackage{authblk}
\usepackage{xifthen}
\usepackage{natbib} 
\usepackage[per-mode=fraction]{siunitx}
\usepackage{amsmath}
\numberwithin{equation}{section}
\usepackage{amsfonts}
\usepackage{mathtools}
\usepackage{bm}
\usepackage{graphicx}
\usepackage[export]{adjustbox}
\usepackage{caption}
\usepackage{subcaption}
\usepackage{chngcntr}
\counterwithin{figure}{section}
\usepackage[colorlinks,linkcolor=blue,citecolor=blue]{hyperref}
\usepackage{csquotes}
\usepackage[nameinlink]{cleveref}
\usepackage{color}

\usepackage{pgfplots}
\pgfplotsset{compat=newest}
\usepackage{tikz}
\usetikzlibrary{3d}
\usetikzlibrary{pgfplots.colorbrewer}
\pgfplotsset{colormap/RdBu}
\pgfplotsset{colormap/PuBu}
\pgfplotsset{colormap/Oranges}

\usepackage{algorithm}
\usepackage{algpseudocode}
\usepackage{multirow}

\usepackage[colorinlistoftodos,prependcaption]{todonotes}

\let\cite\citep

\newcommand{\DAindex}{n}
\newcommand{\DAstep}[1][]{\ifthenelse{\isempty{#1}}{^\DAindex}{^{\DAindex,#1}}}   
\newcommand{\initDAstep}{^0} 
\newcommand{\prevDAstep}[1][]{\ifthenelse{\isempty{#1}}{^{\DAindex-1}}{^{\DAindex-1,#1}}}

\newcommand{\memberIndex}{e}
\newcommand{\member}{_\memberIndex}   

\newcommand{\ensembleSize}{{N_e}}
\newcommand{\obsSize}{{N_Y}}
\newcommand{\stateSize}{{N_X}}
\newcommand{\locStateSize}{{N_{X}^{\text{loc}}}}

\newcommand{\state}{\bm{x}}

\newcommand{\obs}{\bm{y}}

\newcommand{\modelOperator}{\mathcal{M}}
\newcommand{\modelMatrix}{\mathbf{M}}
\newcommand{\modelError}{\bm{\nu}}
\newcommand{\modelErrorCov}{\mathbf{Q}}

\newcommand{\obsMatrix}{\mathbf{H}}
\newcommand{\obsError}{\boldsymbol \epsilon}
\newcommand{\obsErrorCov}{\mathbf{R}}

\newcommand{\group}{\mathcal{B}}
\newcommand{\groupIndex}{b}
\newcommand{\groupSize}{B}
\newcommand{\locWeight}{\mathbf{w}}
\newcommand{\local}[1][]{\ifthenelse{\isempty{#1}}{_\text{loc}}{_{\text{loc},#1}}}

\newcommand{\density}{p}
\newcommand{\proposalDensity}{q}
\newcommand{\mean}{\bm{\mu}}
\newcommand{\CovMatrix}{\bm{\Sigma}}
\newcommand{\weight}{w}
\newcommand{\target}{_\text{target}}

\newcommand{\Cov}{\mathbb{C}\text{ov}}
\newcommand{\Corr}{\mathbb{C}\text{orr}}

\newcommand{\loc}{\bm{s}}
\newcommand{\locIndex}{i}
\newcommand{\gridSize}{N_s}

\newcommand{\simTime}{t}

\newcommand{\R}{\mathbb{R}}
\newcommand{\normal}{\mathcal{N}}
\newcommand{\inv}{^{-1}}
\newcommand{\Id}{\bm{I}}

\definecolor{pt1}{RGB}{187,85,102} 
\definecolor{pt2}{RGB}{0,68,136} 
\definecolor{pt3}{RGB}{204,51,17} 
\definecolor{pt4}{RGB}{0,153,136} 

\definecolor{pt-red}{RGB}{187,85,102} 
\definecolor{pt-blue}{RGB}{0,68,136} 
\definecolor{pt-brightred}{RGB}{204,51,17} 
\definecolor{pt-teal}{RGB}{0,153,136} 

\title{Comparison of Ensemble-Based Data Assimilation Methods for Sparse Oceanographic Data}

\author[1,2,*]{Florian Beiser}
\author[1]{H{\aa}vard Heitlo Holm}
\author[2]{Jo Eidsvik}
\affil[1]{\textit{Mathematics and Cybernetics, SINTEF Digital, Oslo, Norway}}
\affil[2]{\textit{Department of Mathematical Sciences, NTNU, Trondheim, Norway}}
\affil[*]{florian.beiser@sintef.no}

\date{}

\begin{document}

\maketitle

\begin{abstract}
For oceanographic applications, probabilistic forecasts typically have to deal with i) high-dimensional complex models, and ii) very sparse spatial observations. 
In search-and-rescue operations at sea, for instance, the short-term predictions of drift trajectories are essential to efficiently define search areas, but in-situ buoy observations provide only very sparse point measurements, while the mission is ongoing. 
Statistically optimal forecasts, including consistent uncertainty statements, rely on Bayesian methods for data assimilation to make the best out of both the complex mathematical modeling and the sparse spatial data. 

To identify suitable approaches for data assimilation in this context, we discuss localisation strategies and compare two state-of-the-art ensemble-based methods for applications with spatially sparse observations. 
The first method is a version of the ensemble-transform Kalman filter, where we tailor a localisation scheme for sparse point data. 
The second method is the implicit equal-weights particle filter which has recently been tested for related oceanographic applications.

First, we study a linear spatio-temporal model for contaminant advection and diffusion, where the analytical Kalman filter provides a reference. 
Next, we consider a simplified ocean model for sea currents, where we conduct state estimation and predict drift. 
Insight is gained by comparing ensemble-based methods on a number of skill scores including prediction bias and accuracy, distribution coverage, rank histograms, spatial connectivity and drift trajectory forecasts.

\paragraph{Keywords} Spatio-temporal Statistics, Data Assimilation, Sparse Observations, Oceanographic Applications
\end{abstract}

\section{Introduction}
\label{sec:intro}

Data assimilation is essential for obtaining reliable operational oceanographic and atmospheric forecasts. It gives a framework for updating and calibrating numerical models with observations, so that forecasts become more accurate \cite{Evensen2009,Asch2016}. 
There is diversity in types of observations; examples including satellite imagery, radar measurements, weather stations, and ocean buoys. 
Those might represent a wide range of physical quantities and are differently connected to the dynamical model, such that operational prediction systems rely on heavy pre-processing. Together with complex physical simulations, this usually requires large computational resources.

In addition to the periodically updated operational forecasts, one must be able to deliver focused predictions for local and time critical situations such as search-and-rescue operations or contamination incidents at sea \cite{Breivik2013,Rohrs2018}. 
Then one can collect additional drifter-based in-situ observations that are not yet available in the operational data assimilation systems, and run ensembles of local simplified models enabling fast predictions with associated uncertainty quantification. 
The motivation for this work is to investigate efficient data assimilation methods for sparse observations, complementing the traditional operational machinery.

We limit scope to point observations such as information from buoys. 
Albeit highly informative about the ocean state at the buoy locations, they can be many kilometers apart, and spatio-temporal modeling is required to fill in the gaps between the sparse data.
Buoy information is here used i) to constrain an advection-diffusion process for particle concentration \cite{Foss2021}, and ii) to constrain drift trajectories in an ocean model \cite{Holm2020c}. 
Case i) represents a linear system in space-time, and we can study properties of new data assimilation approaches with the optimal Kalman filter (KF) solution. 
Case ii) involves a highly non-linear dynamical model, and we compare approaches via various performance metrics on a synthetic simulation study. The main motivation for our work on ii) is that of short-term improved predictions for search-and-rescue missions by utilisation of very sparse spatial buoy observations.
\citet{Chen2020a} also examine a set of data assimilation methods for such relevant cases of ocean state and drift trajectory forecasting, 
but the characteristics of their model and observation structure is different from ours.

The mathematical representation of for instance ocean currents in its discretised form typically involves a high-dimensional vector-valued variable allocated to spatial location and temporal units, and a highly non-linear model describes its dynamics.
An effective statistical representation of such systems is that of an ensemble, what enables to predict the ocean states with the associated uncertainties. 
From the statistical perspective all data assimilation methods share the aim of representing a conditional (filtering) distribution after seeing the observations.
Versions of the ensemble Kalman filter (EnKF) of \citet{Evensen1994} are widely used in practice and several numerical adaptions to practical problems are known for this method (cf. \Cref{sec:DA}).
The case of sparse observations imposes challenges on efficiency and quality, and our focus will relate to a version of the local ensemble transform Kalman filter (LETKF). 
While particle filters (PFs), see e.g. \citet{VanLeeuwen2009} or \citet{chopin2020introduction}, are less seen in high-dimensional real-world applications since they are prone to degenerate, \citet{Holm2020c} demonstrated a modern PF approach based on the implicit equal-weight particle filter (IEWPF) of \citet{Zhu2016}, that shows promising forecasting results for drift trajectories.

In this paper, we systematically compare the statistical properties of ensemble-based data assimilation methods for sparse observations in practical oceanographic applications.
We present a revised LETKF algorithm that is tailored to applications with sparse point observations, compare this to the state-of-the-art particle filter IEWPF, and study statistical properties of these methods through the two aforementioned cases.


\Cref{sec:DA} expounds state-of-the-art ensemble-based data assimilation techniques and puts the proposed localisation technique into the context of related work.
In \Cref{sec:advectiondiffusion}, we use a dynamical model based on the advection-diffusion equation to verify the relevant ensemble-based filtering methods against an analytical solution.
\Cref{sec:gpuocean} presents the non-linear simplified ocean model for drift trajectory prediction, and we compare the performance of the data assimilation methods. 
Closing remarks are in \Cref{sec:conclusion}.

%
%

\section{Data Assimilation Problem and Ensemble-based Filtering}
\label{sec:DA}

Spatio-temporal quantities are denoted by $\state(t, \loc)$, for time $t>0$ and location $\loc \in \R^{2}$. Upon discretisation of the spatio-temporal domain of interest, they are represented at grid nodes of spatial locations $\left( \loc_\locIndex \right)_{\locIndex=1}^{\gridSize}$ and time steps $\simTime\DAstep$, $\DAindex=1,\dots$.  
The \emph{state vector} at time $\simTime\DAstep$ is denoted $\state\DAstep \in \R^{\stateSize}$. Vectors can hold more than one quantity per location if necessary. In oceanographic application, the dimension $\stateSize$ is usually very high due to large domains and several physical quantities. 

The numerical model is embraced in the \emph{model operator} $\modelOperator\DAstep$. It propagates the state vector from the previous time step $\simTime\prevDAstep$ to the current $\simTime\DAstep$, defining the so-called forecast state. The model usually describes the physics of the ocean. 
To account for uncertainty coming from external factors, unknown model parameters and non-modelled physics, Gaussian model error $\modelError\DAstep \sim \mathcal{N}(0,\modelErrorCov\DAstep)$ is added every time step. It is assumed that the law of the model error is known, and that the error terms are uncorrelated in time. Starting from initial state $\state\initDAstep$ the model evolves as
\begin{equation}
\label{eq:DAmodel}
    \state\DAstep = \modelOperator\DAstep\state\prevDAstep + \modelError\DAstep, \hspace{5mm} \DAindex=1,\dots~.
\end{equation}

The oceanographic state $\state\DAstep$ is often only partially observed and sometimes even indirectly.
The extraction of an \emph{observation} from the true state vector is denoted by the measurement operator $\obsMatrix\DAstep$, where due to the nature of the problems within this work, we impose the restriction that this operator is linear.
Measurement inaccuracies are represented by the addition of a zero-mean Gaussian error $\obsError\DAstep$ with known covariance matrix, such that $\obsError\DAstep\sim\mathcal{N}(0,\obsErrorCov\DAstep)$, independently for each observation time. 
Observations $\obs\DAstep \in \R^{\obsSize}$ are modeled by
\begin{equation}
\label{eq:DAobs}
    \obs\DAstep = \obsMatrix\DAstep\state\DAstep + \obsError\DAstep,
\end{equation} 
and we assume data comes from this observation model employed for the true, but unknown state.
We consider measurement operators $\obsMatrix\DAstep$ that pick a subset of spatial locations, meaning that the matrix consists of one $1$ entry per row and otherwise $0$ entries. The covariance matrix $\obsErrorCov\DAstep$ is assumed to be diagonal, representing conditional independence between the data, given the state variables. 
A characteristic of many oceanographic applications and key assumption in our setting is that in-situ observations are spatially sparse and of low dimension compared to the high-dimensional state vector, i.e.
\begin{equation}
\label{eq:sparseObs}
    \obsSize \ll \stateSize.
\end{equation}

For notational convenience, we neglect the time step superscript in the model and observation operators, as well as state and measurement errors and their associated covariance matrices. For the state vector the time step superscript is always included. 



\emph{Data assimilation} refers to the workflow of sequentially updating the probability density of the state variables as more data gets available. Often, this results in reduced uncertainty, especially near the observation locations.
This kind of data assimilation is formalised in Bayes' rule.
Using the state's density conditioned on all previous observations $\density(\state\DAstep|\obs^{1:n-1})$ as the \emph{forecast} (prediction or \emph{prior}), this rule is used to assimilate the new observation $\obs\DAstep$ and thus provides the \emph{analysis} (filtering or \emph{posterior}) density  $\density(\state\DAstep | \obs^{1:n})$ of the state.
For stochastic processes with a Markov property as implicitly stated in \cref{eq:DAmodel}, this formalism can be applied recursively
\begin{equation}
\label{eq:Bayes}
    \density(\state\DAstep | \obs^{1:n}) 
    \propto \density(\obs\DAstep|\state\DAstep) \density(\state\DAstep|\obs^{1:n-1}), \hspace{3mm} n=1,2,\ldots,
\end{equation}
starting with only prior information at the first time step.

\paragraph{Kalman filtering} 
In case of a linear model $\modelOperator=\modelMatrix$ and an initial Gaussian distribution for the state $\state\initDAstep \sim \mathcal{N}(\mean\initDAstep, \CovMatrix\initDAstep)$, all forecast and analysis distributions remain Gaussian. The data assimilation problem is Gauss-linear and Bayes' formula \eqref{eq:Bayes} has closed form solutions for the mean vectors and covariance matrices. 
The analytical KF computes these in a recursive manner. It follows a two-step procedure at each time step.
Assuming we have an analysis (a) distribution characterised by the mean $\mean\prevDAstep[a]$ and covariance matrix $\CovMatrix\prevDAstep[a]$, the forecast (f) distribution $\state\DAstep|\obs^{1:n-1} \sim \normal(\mean\DAstep[f],\CovMatrix\DAstep[f])$ is obtained by evolving the given moments from the previous time step to the next observation: 
\begin{subequations}
\label{eq:KFforward}
    \begin{align}
        \mean\DAstep[f] & = \modelMatrix \mean\prevDAstep[a] \\
        \CovMatrix\DAstep[f]  & = \modelMatrix \CovMatrix\prevDAstep[a] \modelMatrix^\top + \modelErrorCov.
    \end{align}
\end{subequations}
The analysis distribution $\state\DAstep|\obs^{1:n} \sim \normal(\mean\DAstep[a],\CovMatrix\DAstep[a])$ is achieved by assimilating the latest observation $\obs\DAstep$ via Bayes' rule for the given Gaussian model, resulting in 
\begin{subequations}
\label{eq:KFupdate}
    \begin{align}
        \mean\DAstep[a] & = \mean\DAstep[f] + \mathbf{K}(\obs\DAstep - \obsMatrix\mean\DAstep[f]) \label{eq:KFupdateMean} \\
        \CovMatrix\DAstep[a]  & = \CovMatrix\DAstep[f] - \mathbf{K}\CovMatrix\DAstep[f]\mathbf{K}^\top. \label{eq:KFupdateCov}
    \end{align}
\end{subequations}
Here, $\mathbf{K} = \CovMatrix\DAstep[f] \obsMatrix^\top ( \obsMatrix\CovMatrix\DAstep[f]\obsMatrix^\top + \obsErrorCov)^{-1}$ is the Kalman gain, which in \cref{eq:KFupdateMean} maps the so-called innovation, $\obs\DAstep - \obsMatrix\mean\DAstep[f]$, to state space with respect to the state and observation error covariance matrices. 
From the numerical perspective, note that the Kalman filter requires storage and propagation of the size $\stateSize\times\stateSize$ covariance matrix which is burdensome when the state dimension is large.

\paragraph{Ensemble-based data assimilation} 
In oceanographic applications the linearity assumptions of the Kalman filter are rarely met, and linearised approaches can suffer from divergence challenges. There is hence a need for more flexible methods, and ensemble-based approaches have been employed to gain realism in non-linear systems, while maintaining computational efficiency even for large $\stateSize$. 
Therein, the continuous distribution of the state variable is approximated by an ensemble of realisations $\left(\state\DAstep\member\right)_{\memberIndex=1}^\ensembleSize$ and potentially by corresponding weights $\left( \weight\member \right)_{\memberIndex=1}^\ensembleSize$. Following the Monte Carlo idea, the marginal distribution of $\state\DAstep$ at time $\simTime\DAstep$ becomes
\begin{equation}
\label{eq:ensembleBased}
    \density(\state\DAstep) \approx \frac1\ensembleSize \sum_{\memberIndex=1}^\ensembleSize{ \weight\member\DAstep \delta\left(\state\DAstep - \state\member\DAstep \right)},
\end{equation}
where $\delta$ is the Dirac delta function. 


In the statistical literature, see e.g. \citet{Asch2016} and \citet{Vetra-Carvalho2018}, there are two popular groups of methods for ensemble-based data assimilation, whose foundations and latest variants for the aforementioned problems is outlined in the next sections. 
We focus on representations with equal weights $\weight\member\DAstep=1$, but each of the ensemble members are adjusted in the data assimilation step as well as in the forecast step.


\subsection{Particle filters in oceanographic applications}
\label{sec:pfInOceanApps}

PFs in their simplest forms are ensemble-based methods for solving the data assimilation problem using the Monte Carlo approach. 
Starting from a weighted ensemble approximation for $\density(\state\initDAstep)$ or $\density(\state\prevDAstep|\obs^{1:n-1})$ in the form of \cref{eq:ensembleBased}, the forecast distribution $\density(\state\DAstep|\obs^{1:n-1})$ can be approximated by propagating each ensemble member $\state\member\prevDAstep$ individually by the model using \cref{eq:DAmodel} to get $\state\member\DAstep$. 
Plugging this into Bayes formula \eqref{eq:Bayes}, the new weights are
\begin{equation}
    \weight\member\DAstep \propto \density(\obs\DAstep|\state\member\DAstep) \density(\state\member\DAstep|\state\member\prevDAstep) \weight\member\prevDAstep. 
    \label{eq:pfWeightUpdate}
\end{equation}
Since ensemble members that have weights very close to zero do not contribute to the posterior probability distribution, it is common to combine \cref{eq:pfWeightUpdate} with a discrete resampling of the ensemble members based on their weights (cf. \citet{VanLeeuwen2009,chopin2020introduction} for reviews of resampling schemes). In practice, this means that we discard ensemble members with low weights and duplicate those with higher weights, thus ensuring that computational resources are used to describe the non-negligible part of the distribution. 
In high-dimensional oceanographic application, however, these basic PFs are prone to degenerate, i.e. all but one ensemble member get a weight close to zero, leading to loss of statistical properties~\cite{Snyder2008}.

Among other concepts, one way to counteract such degeneracy is to sample from a proposal density $\proposalDensity\member$ instead of evolving the ensemble directly according to $\density(\state\member\DAstep|\state\member\prevDAstep)$ \cite{VanLeeuwen2019}.
The proposal density can be conditioned on the latest observation $\obs\DAstep$ and the previous state $\state\member\prevDAstep$ for all ensemble members $\memberIndex=1,\dots,\ensembleSize$.
The weights are then modified to
\begin{equation*}
    \weight\DAstep[\ast]\member = \frac{\weight\member\DAstep}{\proposalDensity\member(\state\member\DAstep|\state\prevDAstep_{1:\ensembleSize}, \obs\DAstep)}.
\end{equation*}
The variance in the weights can be reduced in this way, and the minimal variance is achieved by $\proposalDensity\member(\state\DAstep|\state\prevDAstep_{1:\ensembleSize}, \obs\DAstep) = \density(\state\DAstep|\state\prevDAstep\member, \obs\DAstep)$
as described in \citet{Doucet2000} and often referred to as the optimal proposal. 
Note that in our case of Gaussian errors and linear observation operator, this proposal is a Gaussian distribution $\normal\left(\state\DAstep[\text{opt}]\member, \mathbf{P} \right)$ with 
\begin{subequations}
    \begin{align}
        \state\DAstep[\text{opt}]\member &= \modelOperator\state\member\prevDAstep + \modelErrorCov\obsMatrix^\top\left(\obsMatrix\modelErrorCov\obsMatrix^\top + \obsErrorCov\right)\inv \left(\obs\DAstep - \obsMatrix\modelOperator\state\member\prevDAstep\right) \label{eq:optStateUpdate}\\
        \mathbf{P} &=\modelErrorCov- \modelErrorCov\obsMatrix^\top\left(\obsMatrix\modelErrorCov\obsMatrix^\top + \obsErrorCov\right)\inv \obsMatrix\modelErrorCov.
        \label{eq:optCovUpdate}
    \end{align}
\end{subequations}
The expression for $\state\DAstep[\text{opt}]\member$ is similar to the KF update in \cref{eq:KFupdateMean}, but it uses the covariance structure from the model error $\modelErrorCov$ instead of the forecast $\CovMatrix\DAstep[f]$. 
Even this formulation of the PF will degenerate for high-dimensional systems, cf. \citet{Morzfeld2017}.

The optimal proposal density filter can be modified such that all posterior ensemble members obtain a certain target weight $\weight\DAstep\target$.
Instead of drawing realisations from the proposal distribution directly, the IEWPF first samples $\bm{xi}\member$ and $\bm{nu}\member$ from $\normal(0,\Id_\stateSize)$ and next implicitly transforms samples to a target distribution.
This filter, introduced by \citet{Zhu2016} and modified by \citet{Skauvold2019}, utilises a version of the optimal proposal density where $\bm{xi}\member$ and $\bm{nu}\member$ are constructed to be perpendicular and scaled according to factors $\alpha\member^{1/2}$ and $\beta^{1/2}$ before being transformed by $\mathbf{P}$ according to
\begin{equation}
    \state\DAstep\member = \state\DAstep[\text{opt}]\member + \mathbf{P}^{1/2}\left(\alpha\member^{1/2} \bm{xi}\member + \beta^{1/2}\bm{nu}\member\right).
    \label{eq:iewpfUpdate}
\end{equation}
Here, the $\alpha\member$ values are calculated implicitly to ensure equal weights, whereas $\beta$ is a constant tuning parameter influencing the statistical quality of the results.
A small tuning parameter $\beta$ gives small spread of the ensemble that likely underestimates the variability, whereas a bigger $\beta$ increases the spread. \citet[Appendix A]{Holm2020c} derived lower and upper bounds for this tuning parameter. 
In the subsequent experiments, we tune $\beta$ manually, mainly by calibration of coverage probabilities as suggested in \citet{Skauvold2019}.
By experience it seems that values around 0.5 are a good start. Albeit the choice of $\beta$ is independent of the ensemble size, it is influenced by the dynamics of the problem. Hence, one can find a suitable choice of $\beta$ for a specific kind of scenario and then keep it fixed in future experiments with similar characteristics. 

There are no guarantees on how the IEWPF performs, even when the ensemble size goes to infinity. Still, the performance tends to be very good in large-size systems. With $\beta=0$, the implicit transform has a gap that leads to asymptotic bias \citep{Skauvold2019}, but this seems to be adjusted reasonably well by the second part having $\beta>0$.
The IEWPF has recently shown applicable and efficient for assimilating point-based observations into a simplified ocean model based on the shallow water equations~\cite{Holm2020c}. Herein, this method represents a state-of-the-art PF and is investigated more thoroughly.


\subsection{Ensemble Kalman filters in oceanographic applications}
\label{subsec:enkf}

The EnKF~\cite{Evensen1994,Evensen2009} is an ensemble-based version of the KF, given in \cref{eq:KFforward,eq:KFupdate}. Originally presented as a data assimilation method for non-linear systems, it also solves the problem of having to store and propagate the $\stateSize\times\stateSize$ state covariance matrix $\CovMatrix$.

In the ensemble representation from \cref{eq:ensembleBased}, all weights are kept a priori equal to one, and the state of each ensemble member is propagated by the model in \cref{eq:DAmodel}.
The state covariance can then be estimated through the ensemble, as
\begin{equation}
    \widehat{\CovMatrix}\DAstep[f] = \frac{1}{\ensembleSize-1} \sum_{\memberIndex=1}^{\ensembleSize}{\left( \state\member\DAstep[f] - \overline{\state}\DAstep[f] \right) \left( \state\member\DAstep[f] - \overline{\state}\DAstep[f] \right)^\top },
    \label{eq:ensCovEstimate}
\end{equation}
where $\overline{\state}\DAstep[f]$ denotes the ensemble mean.
The ensemble members are then updated along the same linear projection 
\begin{equation}
\label{eq:EnFKupdate}
    \state\member\DAstep[a] = \state\member\DAstep[f] + \widehat{\mathbf{K}}\left( \obs\DAstep - \obsMatrix\state\member\DAstep[f] - \obsError\DAstep_e \right),
\end{equation}
where the Kalman gain becomes $\widehat{\mathbf{K}} = \widehat{\CovMatrix}\DAstep[f]\obsMatrix^\top ( \obsMatrix\widehat{\CovMatrix}\DAstep[f]\obsMatrix^\top + \obsErrorCov)^{-1}$. 
In \cref{eq:EnFKupdate}, the perturbation $\obsError\DAstep\sim\mathcal{N}(0,\obsErrorCov\DAstep)$ is added to adjust the variance in the solution ensemble, motivated by exact sampling in the linear Gaussian situation.
The solution is therefore termed the stochastic EnKF (SEnKF) \cite{Burgers1998, Houtekamer1998, vanLeeuwen2020}.
The SEnKF requires that we obtain and store the relevant covariances from the ensemble, and then factorize matrices to solve the linear system in \cref{eq:EnFKupdate}. 
For high-dimensional applications, this quickly becomes expensive, and it is therefore common to use deterministic square-root formulations instead.

To avoid working in the state space, the ensemble transform Kalman filter (ETKF) reformulates \cref{eq:EnFKupdate} via linear algebraic identities into so-called deterministic square-root formulations~\cite{Whitaker2002}, which works in ensemble dimensions instead.
Mathematically, let $\mathbf{X}\DAstep[f] = \left[\state_1\DAstep[f], \dots, \state_{\ensembleSize}\DAstep[f] \right]$ be the matrix of prior ensemble states, and let $\overline{\mathbf{X}}\DAstep[f]$ be a $\stateSize\times\ensembleSize$ matrix where all columns are $\overline{\state}\DAstep[f]$.
ETKF then works on the state perturbation matrix $\mathbf{X}\DAstep[f]_\text{pert} = \mathbf{X}\DAstep[f] - \overline{\mathbf{X}}\DAstep[f]$, 
and calculates the mean of the analysis ensemble \begin{equation}
   \overline{\mathbf{X}}\DAstep[a] = \overline{\mathbf{X}}\DAstep[f] +  \mathbf{X}\DAstep[f]_\text{pert}\mathbf{A}(\mathbf{H} \mathbf{X}_\text{pert}^{n,f})^{\top}\obsErrorCov^{-1}\left( \obs\DAstep - \obsMatrix \overline{\state}\DAstep[f] \right),
\end{equation}
where
\begin{equation}
    \mathbf{A} = \left(  (\ensembleSize-1) \mathbf{I}_{\ensembleSize} + (\mathbf{H} \mathbf{X}_\text{pert}^{n,f})^{\top} \obsErrorCov \mathbf{H} \mathbf{X}\DAstep[f]_\text{pert} \right)^{-1}
\end{equation}
plays the role of the analysis covariance matrix. 
The ensemble members are then spread around $\overline{\state}\DAstep[a]$ according to
\begin{equation}
    \mathbf{X}\DAstep[a] = \overline{\mathbf{X}}\DAstep[a]  + \mathbf{X}\DAstep[f]_\text{pert} \left( (\ensembleSize-1) \mathbf{A} \right)^\frac{1}{2},
\end{equation}
where we use a singular-value decomposition to find the square-root of $\mathbf{A}$.
The properties of the ETKF remain the same as for the EnKF and we refer to \citet{Li2007} for further details on the transform.

The derivation of these methods assume a linear model, and asymptotic convergence results for increased ensemble size cannot be proved for non-linear cases. Still, the EnKF and its variants have been prevalent and successfully used in oceanographic applications, cf. \citet{Carrassi2018}. 

The error covariance matrix in this kind of filters is estimated from the ensemble and can lead to systematic underestimation.
Typical approaches to counteract this are inflation or localisation. 
\citet{Anderson1999} introduced \emph{covariance inflation} by a multiplicative factor to keep more variability in the ensemble, where several suggestions for the determination of an adaptive factor exist in literature, see e.g. \citet{Desroziers2006, Anderson2009, Saetrom2013, Raanes2019}. 
However, \citet{Li2009SimultaneousFilter} also point out that covariance inflation may not work appropriately in large complex models.
Hence, we will mainly concentrate on localisation. 


\subsection{Sparse observations} 

The focus of this paper is on assimilating spatially sparse point observations, which naturally suggests \emph{localisation}. Although localisation is important also for general applications, the sparseness considered here motivates one to study specialised methods in terms of assimilation quality and algorithmic efficiency.
%


\paragraph{Localisation and sparse observation handling in the IEWPF}
The need for localisation in EnKF-based schemes arises from the spurious correlations introduced by the term $\widehat{\CovMatrix}\DAstep[f] \obsMatrix^{\top}$ representing estimated correlations between all observations and all variables through the ensemble in \cref{eq:EnFKupdate}. As pointed out in Section~\ref{sec:pfInOceanApps}, the optimal proposal in \cref{eq:optStateUpdate} updates the state vector with a similar expression, but it uses the model correlations in $\modelErrorCov \obsMatrix^{\top}$ rather than the empirical $\widehat{\CovMatrix}\DAstep[f] \obsMatrix^{\top}$. This means that the optimal proposal filter does not lead to spurious correlations. Furthermore, if $\modelErrorCov$ is local and do not overlap between observation sites, the updates in \cref{eq:optStateUpdate} are also local.

For this to apply in the IEWPF update in \cref{eq:iewpfUpdate}, note that \cref{eq:optCovUpdate} can be written as
\begin{equation}
        \mathbf{P} = \modelErrorCov^{1/2 \top} \Big( \Id_\stateSize - \underbrace{\modelErrorCov^{1/2} \obsMatrix^{\top} ( \obsMatrix \modelErrorCov \obsMatrix^{\top} + \obsErrorCov)^{-1} \obsMatrix \modelErrorCov^{1/2 \top}}_{=:\mathbf{S}} \Big) \modelErrorCov^{1/2}. \\
    \label{Pidentity} 
\end{equation}
By considering an observation of a single state variable only, we see that $(\obsMatrix \modelErrorCov \obsMatrix^{\top} + \obsErrorCov)^{-1}$ becomes a scalar, and $\obsMatrix^{\top} (\cdot) \obsMatrix$ maps it onto the state space, leading to a $\stateSize \times \stateSize$ matrix consisting of only a single non-zero value. This value is then spread in the state space according to the model error covariance matrix as $\modelErrorCov^{1/2} (\cdot) \modelErrorCov^{1/2 \top}$. Hence, assuming that the model error has a local covariance structure,  $\mathbf{S}$ has non-zero values only in rows and columns that are local to the observation location. Far from the observations, $\mathbf{P}$ is  identical to $\modelErrorCov$.
In \cref{eq:iewpfUpdate}, $\bm{xi}\member$ and $\bm{nu}\member$ are perpendicular random Gaussian distributed vectors, and $\alpha\member$ and $\beta$ are scalars. This means that information spread by $\mathbf{P}^{1/2}$ in \cref{eq:iewpfUpdate} is similar to the model error term far from observations, and slightly modified due to $\mathbf{S}$ near observations.

This shows that the IEWPF has built-in localisation as long as the covariance structure of the model error is local.
It should be noted though, that the values of $\alpha\member$ and $\beta$ depend on the innovation from all observations in the domain.
These parameters are therefore global parameters, but since they are scalars, they do not contribute to any spurious correlations. 

Regarding efficient handling of sparse observations, we first observe from \cref{eq:optStateUpdate} that $\state\DAstep[\text{opt}]\member$ can be constructed independently for each ensemble member.
Further, assuming that the correlation structures in $\mathbf{Q}$ does not overlap between observation sites, the contribution to $\state\DAstep[\text{opt}]\member$ from each observation can also be evaluated independently.
We do, however, need to synchronise all ensemble members to gather all innovation vectors within the ensemble to compute $\beta$, but this is a cheap operation with sparse observations.
To apply $\mathbf{P}^{1/2}$ next, we can once again evaluate the contribution from $\mathbf{S}$ independently for each observation under the same sparsity assumption.
(See also \citet{Holm2020}).


\paragraph{Localisation in the EnKF}
In the statistical sense, the spurious correlations in the EnKF are due to a poor Monte Carlo approximation of the true covariance matrix, cf. \citet{Houtekamer2016}. In the spatio-temporal physical perspective one can say that they contradict the finite speed of information propagation. 
Prevailing techniques to counteract these artefacts are covariance or observation localisation as they are outlined in \citet{Sakov2011}. Both of these exploit the physical distance between two points in space to reduce information propagation effects, and this has been demonstrated to work well in practice, see e.g. \citet{Soares2021}. 
For many oceanographic applications, it is important that the geostrophic imbalance introduced by the localisation in the EnKF does not outweigh the natural imbalance - \citet{Greybush2011} provide a discussion and representative experiments to this issue.
\citet{Ott2004} introduce an efficient localisation scheme for the ETKF relying on the local areas. This is popular in numerical weather prediction, cf. \citet{Szunyogh2007}. 
The resulting local (LETKF) follows the paradigm: 
\begin{displayquote}
    The analysis in a local area is only influenced by the observations in its neighborhood.
\end{displayquote}
This means that one loops over the state locations or set of state locations in a batch area, and updates those by means of the ETKF using a specified set of observations per batch.
Assuming that the data are conditionally independent, one does not need to modify the data error covariance as batches of data are assimilated.

\paragraph{Sparse observation handling in the ETKF}
In an oceanographic scenario with observations at only a few locations, the typical localisation in the LETKF meets characteristic challenges.
On one hand, the definition of a reasonably small local area gets computationally prohibitively expensive due to the loop over all of them. On the other side, many areas only assimilate a single observation or become reset to their mean. 
More efficient implementations using parallelisation and observation batching exist, see e.g. \citet{Hunt2007}. This works well for applications in numerical weather prediction, but for the very little number of observation locations that we tend to have with buoys in the ocean the schemes still struggles, and none of them exploit this sparsity of the observations explicitly. 

We propose a scheme tailored to the sparsity of observations, changing the paradigm:
\begin{displayquote}
    A certain observation influences only the analyses of the spatial variables within its neighbourhood.
\end{displayquote}
This yields the definition of local domains around each observation site only, where the size $\locStateSize$ of a local area is significantly smaller than the full state space. 
For the choice of the radius of the resulting local domains, several approaches exist, see e.g. \citet{Kirchgessner2014}, but we advocate using model-informed radii like the model error range if possible. 
Then, we assume that observations with non-overlapping local areas are nearly uncorrelated, such that their potential correlation can be neglected. 
Due to the motivating postulate, we expect to define less areas than in the traditional approach that only overlap very seldom.

\begin{figure}[ht]
\hspace*{-1cm}
\begin{subfigure}[b]{0.2\textwidth}
\begin{tikzpicture}[scale=0.75]
    \begin{axis}[xmin=0,xmax=10, ymin=0,ymax=10, zmin=0,zmax=1, hide axis]
        \def\x{7.0}
        \def\y{3.0}
	\def\xx{2.0}
	\def\yy{4.0}
	\def\xxx{2.5}
	\def\yyy{4.5}
	\def\xxxx{4.0}
	\def\yyyy{8.0}
	\def\xxxxx{5.5}
	\def\yyyyy{7.5}
        \begin{scope}[canvas is xy plane at z=0]
            \draw[step=5mm, black, line width=0mm] (0,0) grid (10,10);
            \draw[black, ultra thick] (\x-1,\y-1) rectangle (\x+1.5,\y+1.5);
            \fill[black] (\x,\y) rectangle (\x+0.5,\y+0.5);
            \draw[black, ultra thick] (\xx-1,\yy-1) rectangle (\xx+1.5,\yy+1.5);
            \fill[black] (\xx,\yy) rectangle (\xx+0.5,\yy+0.5);
            \draw[black, ultra thick] (\xxx-1,\yyy-1) rectangle (\xxx+1.5,\yyy+1.5);
            \fill[black] (\xxx,\yyy) rectangle (\xxx+0.5,\yyy+0.5);
            \draw[black, ultra thick] (\xxxx-1,\yyyy-1) rectangle (\xxxx+1.5,\yyyy+1.5);
            \fill[black] (\xxxx,\yyyy) rectangle (\xxxx+0.5,\yyyy+0.5);
            \draw[black, ultra thick] (\xxxxx-1,\yyyyy-1) rectangle (\xxxxx+1.5,\yyyyy+1.5);
            \fill[black] (\xxxxx,\yyyyy) rectangle (\xxxxx+0.5,\yyyyy+0.5);
        \end{scope}
    \end{axis}
\end{tikzpicture}
\end{subfigure}
\hspace{2.5cm}
\begin{subfigure}[b]{0.2\textwidth}
\begin{tikzpicture}[scale=0.75]

    \begin{axis}[xmin=0,xmax=10, ymin=0,ymax=10, zmin=0,zmax=2, hide axis]
        \def\x{7.0}
        \def\y{3.0}
	\def\xx{2.0}
	\def\yy{4.0}
	\def\xxx{2.5}
	\def\yyy{4.5}
	\def\xxxx{4.0}
	\def\yyyy{8.0}
	\def\xxxxx{5.5}
	\def\yyyyy{7.5}
        \begin{scope}[canvas is xy plane at z=1]
            \draw[step=5mm, black!25] (0,0) grid (10,10);
            \draw[pt1, ultra thick] (\x-1,\y-1) rectangle (\x+1.5,\y+1.5);
            \fill[pt1] (\x,\y) rectangle (\x+0.5,\y+0.5);
            \draw[step=5mm, pt1] (\x-1,\y-1) grid (\x+1.5,\y+1.5);
        \end{scope}
        \begin{scope}[canvas is xy plane at z=0]
            \draw[step=5mm, black!25] (0,0) grid (10,10);
            \draw[pt1, ultra thick] (\xx-1,\yy-1) rectangle (\xx+1.5,\yy+1.5);
            \fill[pt1] (\xx,\yy) rectangle (\xx+0.5,\yy+0.5);
            \draw[step=5mm, pt1] (\xx-1,\yy-1) grid (\xx+1.5,\yy+1.5);
        \end{scope}
        \begin{scope}[canvas is xy plane at z=2]
            \draw[step=5mm, black!25] (0,0) grid (10,10);
            \draw[pt1, ultra thick] (\xxxx-1,\yyyy-1) rectangle (\xxxx+1.5,\yyyy+1.5);
            \fill[pt1] (\xxxx,\yyyy) rectangle (\xxxx+0.5,\yyyy+0.5);
            \draw[step=5mm, pt1] (\xxxx-1,\yyyy-1) grid (\xxxx+1.5,\yyyy+1.5);
        \end{scope}
    \end{axis}
    	
\end{tikzpicture}

\end{subfigure}
\hspace{2.5cm}
\begin{subfigure}[b]{0.2\textwidth}
\begin{tikzpicture}[scale=0.75]
    
    \begin{axis}[xmin=0,xmax=10, ymin=0,ymax=10, zmin=0,zmax=2, hide axis]
        \def\x{7.0}
        \def\y{3.0}
	\def\xx{2.0}
	\def\yy{4.0}
	\def\xxx{2.5}
	\def\yyy{4.5}
	\def\xxxx{4.0}
	\def\yyyy{8.0}
	\def\xxxxx{5.5}
	\def\yyyyy{7.5}
        \begin{scope}[canvas is xy plane at z=0]
            \draw[step=5mm, black!25] (0,0) grid (10,10);
            \draw[blue, ultra thick] (\xxx-1,\yyy-1) rectangle (\xxx+1.5,\yyy+1.5);
            \fill[blue] (\xxx,\yyy) rectangle (\xxx+0.5,\yyy+0.5);
            \draw[step=5mm, blue] (\xxx-1,\yyy-1) grid (\xxx+1.5,\yyy+1.5);
        \end{scope}
        \begin{scope}[canvas is xy plane at z=1]
            \draw[step=5mm, black!25] (0,0) grid (10,10);
            \draw[blue, ultra thick] (\xxxxx-1,\yyyyy-1) rectangle (\xxxxx+1.5,\yyyyy+1.5);
            \fill[blue] (\xxxxx,\yyyyy) rectangle (\xxxxx+0.5,\yyyyy+0.5);
            \draw[step=5mm, blue] (\xxxxx-1,\yyyyy-1) grid (\xxxxx+1.5,\yyyyy+1.5);
        \end{scope}
	\begin{scope}[canvas is xy plane at z=2]
                \draw[step=5mm, black!0] (0,10) grid (10,10);
        \end{scope}
    \end{axis}
    	
\end{tikzpicture}
\end{subfigure}
\caption{Schematic decomposition of the physical space into local areas around observations (indicated by a filled grid cell) and the separation into groups of uncorrelated observation: In the leftmost grid, five observation sites are marked and the local areas are defined around. Four out of the five local areas intersect with another, such that the observations are split into two groups of uncorrelated data. The observations and local areas in group 1 are signified in red in the middle column and the ones of group 2 are drawn in blue in the right column.}
\label{fig:localareas}
\end{figure}
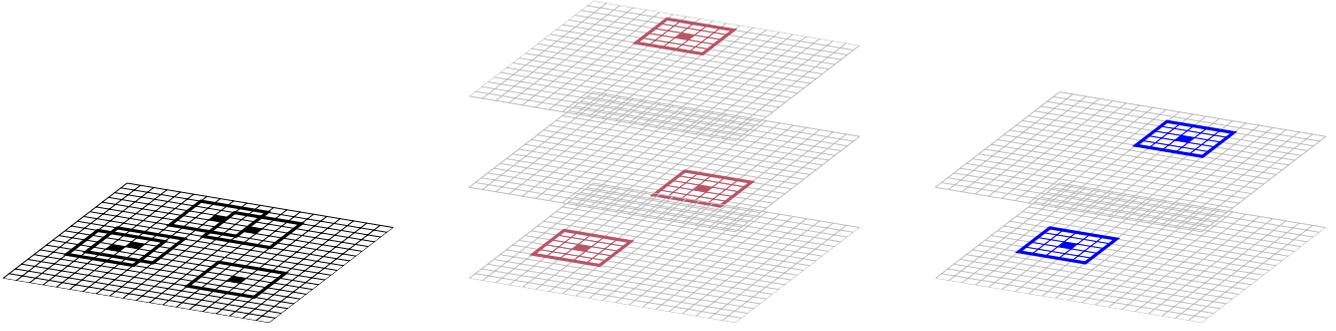

With sparse spatial observations, the correlation in the data is typically rather small, and this means that local computations are efficient. Still, with non-linear dynamical models, it is sometimes difficult to predict the effect of local approximations. Using sequential data integration, one can run through the data in multiple assimilation steps, properly accounting for the correlations. 
In cases of overlapping local observation areas, we therefore recommend splitting the observations into observational batches $\group_\groupIndex, \groupIndex=1,\dots,\groupSize$ of assumed uncorrelated observations for serial processing as originally introduced in \citet{Houtekamer1998}.
In our context, the batches at each step are constructed from far-apart observations. \citet{Nerger2015} discusses that interactions of localisation and serial observation processing could destabilise the filter, but in realistic large-scale applications as we deal with this is not significant.
\Cref{fig:localareas} illustrates the definition of local areas around depicted observation sites, and the subsequent splitting into batches such that all observations sites within a batch are sufficiently far away from each other.

Within the recursion of batches $\group_\groupIndex$, $\state\DAstep[a](j)\in\locStateSize$ are calculated independently in each local area around the observation sites $j\in\group_\groupIndex$.
Note that the processing of a batch can influence the ensemble that is used as prior for the next. 
The analyses only take values in their respective small local region, but to avoid cumbersome notation for transformations, we abuse the same notation for their extension to the full state size $\state\DAstep[a](j)\in\stateSize$. 
From the resulting local analyses, a global analysis state is constructed such that, at an observation site the most analysis information is used, and far away from any observation site the forecast with its full spread is retained. 
This is realised by weighting the available information
\begin{equation}
\label{eq:weighting}
    \state\DAstep[a,\groupIndex]\member = (1-\locWeight_\groupIndex) \state\DAstep[a,\groupIndex-1]\member + \locWeight_\groupIndex \sum_{j\in\group_\groupIndex}{\state\DAstep[a]\member(j)},
\end{equation}
where $\locWeight_\groupIndex$ are weighting vectors and $\state\DAstep[a,0] = \state\DAstep[f]$.

The weighting should assure that $\locWeight=1$ at an observation site and $\locWeight=0$ outside the local areas. 
While veering away from an observation site, $\locWeight$ should transit decreasingly monotone and smoothly from one to zero. 
An example of a kernel fulfilling those requirements locally in continuous space is the Gaspari-Cohn function (GC) introduced in \citet{Gaspari1999}, which enjoys popularity in EnKF-localisation. 
We let $\locWeight_{\text{loc},j}\in\locStateSize$ be a properly scaled discretisation of the GC kernel around observation $j$, such that its support matches the radius of the local area.
With the same notation, the weighting is composed as 
\begin{equation}
    \locWeight_\groupIndex = \sum_{j\in\group_\groupIndex}{\locWeight_{\text{loc},j}}.
\end{equation}

\begin{figure}[htb]
    \centering
    \adjustimage{width=0.4\textwidth,Clip=0 0 0.75cm 0}{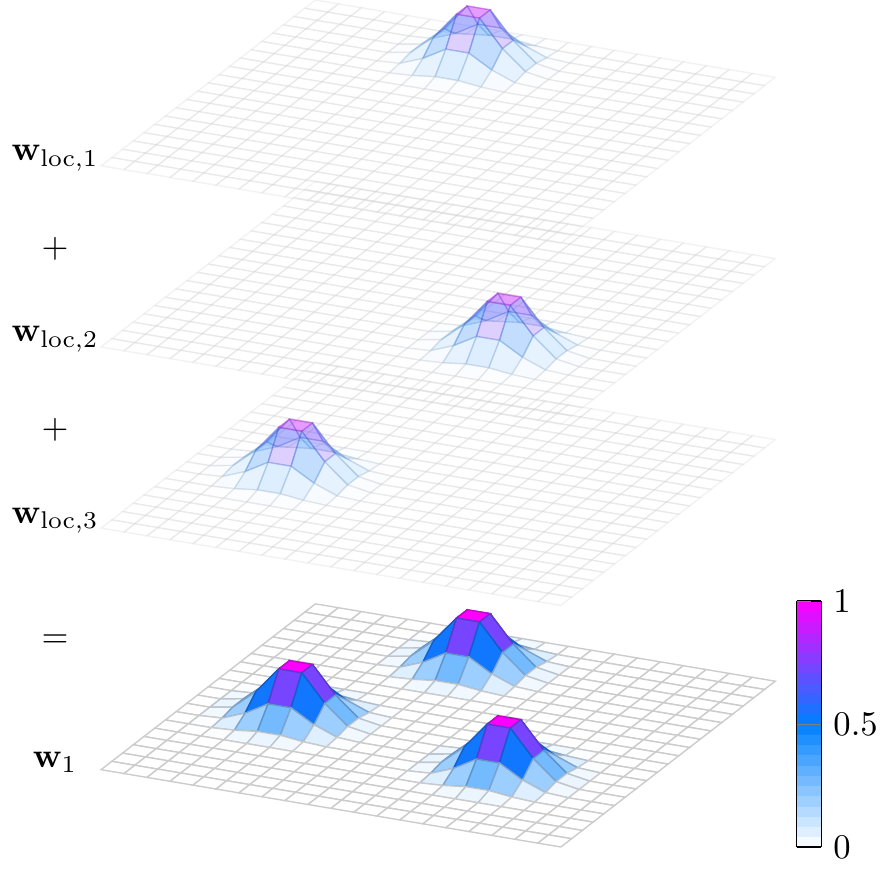}
    \hspace*{1cm}
    \includegraphics[width=0.4\textwidth]{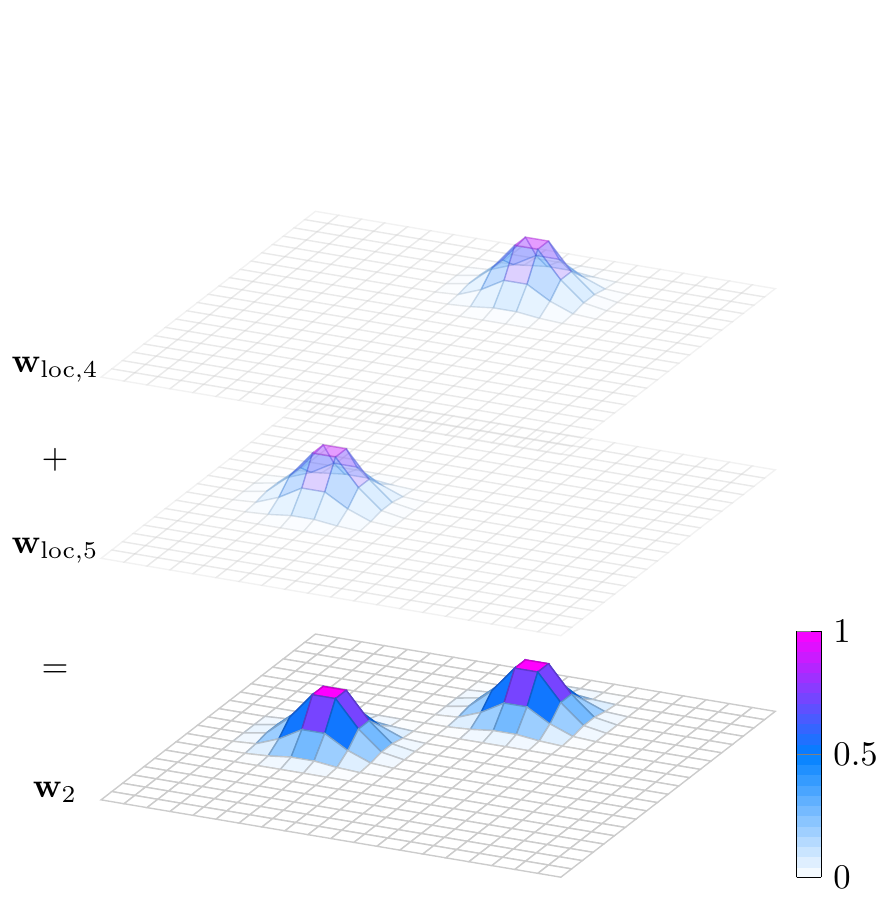}
    \caption{Schematic construction of the weighting $\locWeight_\groupIndex$ using the $\locWeight_{\text{loc},j}$ within the groups: For batch 1 and 2 from \Cref{fig:localareas}, the weighting vectors $\locWeight_1$ and $\locWeight_2$ are built from the Gaspari-Cohn kernel around each observation with in the two groups. At the corresponding center, each $\locWeight\local[j]$ equals one and decays towards the boundary of the local domains. By definition the supports are non-overlapping and the contributions are added up.}
    \label{fig:localweights}
\end{figure}

For the observation batch from \Cref{fig:localareas}, the weighting vectors $\locWeight_1$ and $\locWeight_2$ are illustrated in \Cref{fig:localweights} together with the contributions from $\locWeight\local[1],\dots,\locWeight\local[5]$. 
After the last batch, we set $\state\DAstep[a] = \state\DAstep[a,\groupSize]$ to obtain a final analysis state.

As mentioned in \Cref{subsec:enkf}, covariance inflation is a common but potentially troublesome approach to counteract overfitting and to keep more prior spread in the analysis. 
Here, the presented observation processing offers a neat way to give more weight to the forecast and thereby increase the variance, while avoiding known issues of traditional inflation.
Introducing a scaling parameter $\phi\in[0,1]$ to the weighting vector can alternatively be constructed as 
\begin{equation}
    \locWeight_\groupIndex^{\text{infl}} = \sum_{j\in\group_\groupIndex}{\phi~\locWeight_{\text{loc},j}},
\end{equation}
such that for $\phi=0$ represents a pure Monte-Carlo simulation and $\phi=1$ is the already presented scheme without inflation.

\begin{algorithm}[H]
    \begin{algorithmic}
        \State Given $\mathbf{X}\DAstep[f]$. Parameters: localisation radius and inflation $\phi$ 
        \State Set $\mathbf{X}\DAstep[a,0] = \mathbf{X}\DAstep[f]$
        \For {$\groupIndex=1,\dots,\groupSize$}
            \State Allocate $\locWeight_\groupIndex^{\text{infl}}$ \Comment{\parbox{2cm}{$\stateSize$}}
            \For {$j\in\group_\groupIndex$}
                \State Calculate local $\mathbf{X}\DAstep[a](j)$ using ETKF where using $\mathbf{X}\DAstep[a,\groupIndex-1]$ as forecast in the ETKF
                 \Comment{\parbox{2cm}{$\locStateSize$}}
                \State $\locWeight_\groupIndex^{\text{infl}} \mathrel{{+}{=}} \phi \locWeight_{\text{loc},j}$ 
            \EndFor
            \State $\mathbf{X}\DAstep[a,\groupIndex] = (1-\locWeight_\groupIndex^{\text{infl}}) \mathbf{X}\DAstep[a,\groupIndex-1] + \locWeight_\groupIndex^{\text{infl}} \sum_{j\in\group_\groupIndex}{\mathbf{X}\DAstep[a](j)}$ \Comment{\parbox{2cm}{$\stateSize$}}
         \EndFor
         \State $\mathbf{X}\DAstep[a] = \mathbf{X}\DAstep[a,\groupSize]$
    \end{algorithmic}
    \caption{Analysis scheme using the localisation technique for sparse observation}
    \label{alg:LETKF-analysis}
\end{algorithm}

From a sound statistical perspective, one could process each individual observation in a serial manner, but the collection in prescribed batches reduces iterations. 
In doing so, one assimilates the spatial data recursively, similar to the assimilation over time, albeit without the dynamical state evolution because all updates happen at the time when the data is available. For computational efficiency one again imposes some kind of local routine, and in practice this may rely on GC tapering of the matrices involved.
This is commonly done in implementations of Kriging or applications with sequential uncertainty reduction, where the analysis can depend on the choice of conditioning order, cf. \citet{Nussbaumer2018}.

\section{Comparison against the Analytical Kalman Filter in a Linear Gaussian Advection Diffusion Model}
\label{sec:advectiondiffusion}

In this section, we examine a linear Gaussian spatio-temporal model.
As mentioned in \Cref{sec:DA}, this means that the KF in \cref{eq:KFforward,eq:KFupdate} defines the analytical solution. Ensemble-based approximations and localisation effects of the different filtering techniques from \Cref{sec:DA} can be verified against the KF.

\subsection{Advection diffusion model}

Inspired by \citet{Sigrist2015}, we consider a stochastic advection diffusion equation for state $c$ given by
\begin{equation}
\label{eq:advectiondiffusion}
    \frac{\partial c(\simTime,\loc)}{\partial \simTime} = \nabla \cdot d \nabla c(\simTime,\loc) - \mathbf{v}_\simTime \cdot \nabla c(\simTime,\loc) + \zeta c(\simTime,\loc)  + w(\simTime, \loc).
\end{equation}
The model's parameters are $d=0.25$ for the diffusion, $\mathbf{v} = (1.0,0.1)^\top$ for the advection, and $\zeta=-0.0001$ for the damping. We assume the stochastic error process $w$ has uncorrelated elements over time but smooth dependent spatial components at each time and \eqref{eq:advectiondiffusion} holds for one sampled path of $w$.
We consider a rectangular spatial domain $[0,5] \times [0,3]$ with periodic boundary conditions, and $c$ will be initialised at time $\simTime=0$ as Gaussian random field with Mat\'ern covariance kernel. 

Equation~\eqref{eq:advectiondiffusion} can be used to represent for instance marine pollution dynamics \cite{Foss2021}, 
where the goal is to predict the concentration $c=c_\simTime(\simTime,\loc)$ of a contaminant over time and space in the ocean.
In that case, the advection parameter $\mathbf{v}$ would typically come from a full ocean model if vertical currents are ignored. 


In the discretised setting, the spatial domain is covered by a uniform Cartesian grid with center points $(\loc_i)_{i=1}^{\gridSize}$ in quadratic cells of size $0.1\times0.1$.
The state vector $\state\DAstep$  collects all concentrations $c(\simTime\DAstep,\loc_i)$ at regular time steps $\simTime\DAstep$.
The initial state is represented by a Gaussian random field so that $\state\initDAstep\sim\normal(\mean\initDAstep, \CovMatrix\initDAstep)$ with mean $\mean\initDAstep$ and covariance $\CovMatrix\initDAstep$ having Mat\'ern-type 
\begin{equation*}
    \CovMatrix\initDAstep_{k,l} = \sigma^2 (1+\psi D_{k,l})\exp(-\psi D_{k,l})
\end{equation*}
where $\sigma=0.5$ is the standard deviation at $\loc_k$ and $\psi=3.5$ is the Mat\'ern correlation decay parameter, and 
$D_{k,l}$ is the distance between $\loc_k$ and $\loc_l$. 
The mean $\mean\initDAstep$ is ten in the north-east with higher bell-shaped concentration values in the south-west, see \Cref{fig:advecdiff-kf} (left).

For the numerical solution of the SPDE in \eqref{eq:advectiondiffusion}, a temporal forward and spatial central finite-difference scheme is employed such that the model resembles \cref{eq:DAmodel} with the linear operator $\modelOperator = \modelMatrix$. With periodic boundary condition the low-concentration area leaves the domain on the east boundary and enters from the west. 
The model error $\modelError$ is again represented by a Gaussian random field with a covariance matrix $\modelErrorCov$ of a similar Mat\'ern-type. A smaller standard deviation $\sigma=0.125$ is used, and larger correlation decay parameter $\psi=7.0$ leading to model noise with smaller correlation.

\subsection{Experiment design and analytic solution}

\begin{figure}[h!tb]
    \centering
    \hspace*{-0.5cm}
    \begin{subfigure}{0.25\textwidth}
    \includegraphics[]{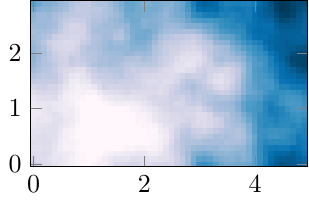}
    \caption*{\quad $t=0$}
    \end{subfigure}
    \hfill
    \begin{subfigure}{0.25\textwidth}
    \includegraphics[]{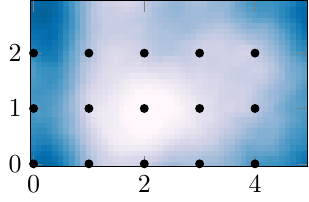}
    \caption*{\quad $t=50$}
    \end{subfigure}
    \hfill
    \begin{subfigure}{0.25\textwidth}
    \includegraphics[]{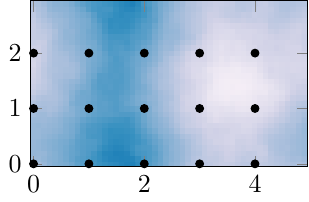}
    \caption*{\quad $t=150$}
    \end{subfigure}
    \hfill
    \begin{subfigure}{0.25\textwidth}
    \includegraphics[]{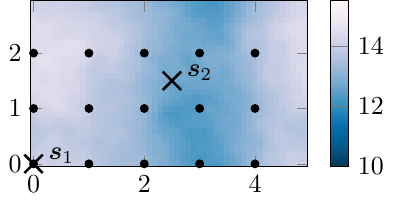}
    \caption*{\quad $t=250$}
    \end{subfigure}
    \caption{The "truth" realisation of the stochastic process at the initialisation and selected observation times with black dots marking the observation sites and black crosses signifying two selected locations of interest.}
    \label{fig:advecdiff-truth}
\end{figure}

A single realisation of the advection diffusion generated by the forward model is used to retrieve observations for the filtering, see \Cref{fig:advecdiff-truth}. It is simulated for 250 time steps with $\Delta t = 0.01$ on a grid of size $50\times30$. The simulated process is observed at $\simTime\DAstep=25n, n=1,\dots,10$ at 15 grid cells marked red in \Cref{fig:advecdiff-truth}. 
These direct state observations are made with a small observation error $\obsError\DAstep\sim\mathcal{N}(0,r^2\mathbf{I})$, $r=0.1$.

\begin{figure}[h!tb]
    \centering
    \hspace*{-0.5cm}
    \begin{subfigure}{0.25\textwidth}
    \includegraphics[]{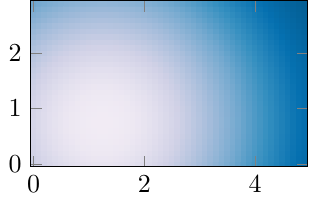}
    \includegraphics[]{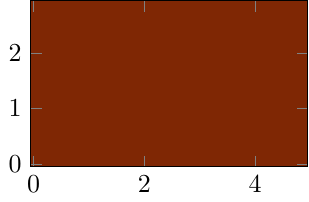}
    \caption*{\quad $t=0$}
    \end{subfigure}
    \hfill
    \begin{subfigure}{0.25\textwidth}
    \includegraphics[]{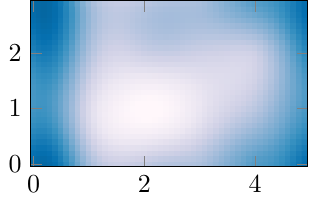}
    \includegraphics[]{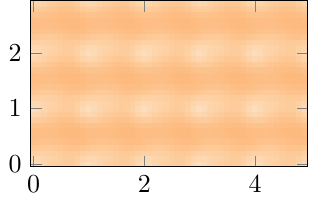}
    \caption*{\quad $t=50$}
    \end{subfigure}
    \hfill
    \begin{subfigure}{0.25\textwidth}
    \includegraphics[]{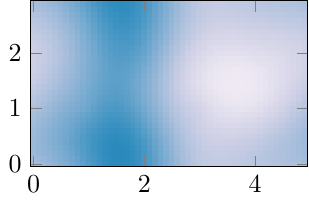}
    \includegraphics[]{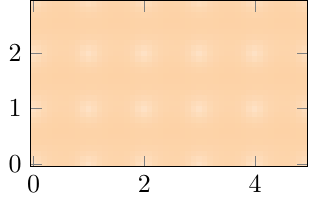}
    \caption*{\quad $t=150$}
    \end{subfigure}
    \hfill
    \begin{subfigure}{0.25\textwidth}
    \includegraphics[]{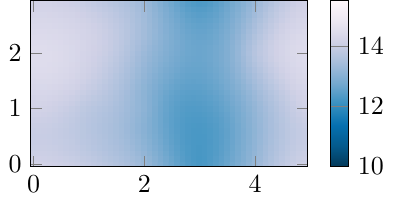}
    \includegraphics[]{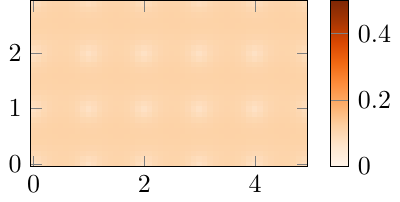}
    \caption*{\quad $t=250$}
    \end{subfigure}
    \caption{For the same times as in \Cref{fig:advecdiff-truth}, the resulting analysis mean (upper row) and standard deviation (lower row) of the KF.}
    \label{fig:advecdiff-kf}
\end{figure}

The KF reference solution is depicted for a selection of time steps in Figure \ref{fig:advecdiff-kf}. 
As for the truth in \Cref{fig:advecdiff-truth}, the filtering mean (\Cref{fig:advecdiff-kf}, top) shows an east north-east movement of the concentrations as expected from the advection term. 
The standard deviations (\Cref{fig:advecdiff-kf}, bottom) are clearly reduced by the data assimilation, especially around the observation sites and in the advection direction.
With time, however, the accuracy of the solution converges as the corrections from doing data assimilation are balanced out by the dynamic model errors.

\subsection{Numerical results and evaluation metrics}

We now solve this concentration advection diffusion problem using the ensemble-based methods IEWPF, ETKF, and LETKF from \Cref{sec:DA}. 
The parameter $\beta$ emerging in the IEWPF is tuned manually and set to 0.55, and this will be discussed further in relation to some of the results. 
We set the localisation radius of the LETKF equal to the correlation range of the model error, which leads to four observational batches. First, we do not use any inflation in the perfect linear model, as suggested by \citet{Raanes2019}, but we return to this below.

The performance of ensemble-based solutions are opposed to the KF reference solution.
We use a set of metrics to evaluate different statistical aspects of the data assimilation methods. 



\paragraph{Root-mean-squared error}
The ensemble mean $\overline{\state}^a$ is compared with the KF mean $\mean^a$.
Here, we consider the state at $t=250$ after assimilating all available observations. 
The error in the mean at each position is then the vector $\textbf{err}_\text{mean}^{\text{KF}} = \left( \mean^a - \overline{\state}^a \right)$.
As a scalar metric to compute the behaviour over all grid cells, we use the root-mean-squared error (RMSE)
\begin{equation}
    \text{RMSE} = \| \textbf{err}_\text{mean}^{\text{KF}} \|_2.
\end{equation}

\begin{figure}[h!tb]
    \centering
    \hspace*{-0.5cm}
    \begin{subfigure}{0.3\textwidth}
    \includegraphics[]{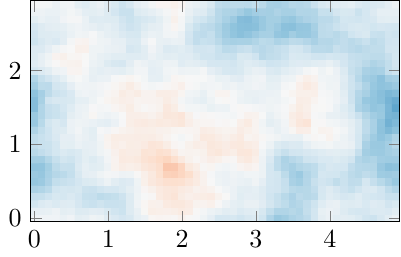}
    \caption{\quad IEWPF $[1.76]$}
    \end{subfigure}
    \hfill
    \begin{subfigure}{0.3\textwidth}
    \includegraphics[]{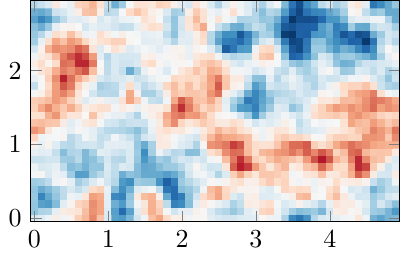}
    \caption{\quad ETKF $[3.35]$}
    \end{subfigure}
    \hfill
    \begin{subfigure}{0.3\textwidth}
    \includegraphics[]{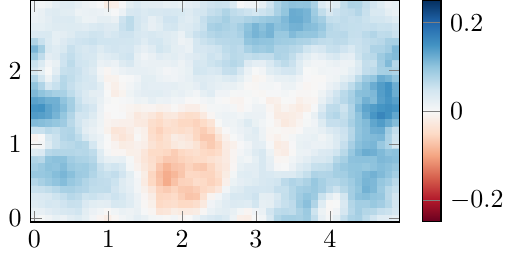}
    \caption{\quad LETKF $[2.26]$}
    \end{subfigure}
    \caption{Mean error $ \textbf{err}_\text{mean}^{\text{KF}}$ at $\simTime=250$ for assimilation experiments with $\ensembleSize=50$ ensemble members. RMSE is given in the bracket for each specified method.}
    \label{fig:advecdiff-rmse}
\end{figure}

\Cref{fig:advecdiff-rmse} shows $\textbf{err}_\text{mean}$ at each grid cell for a single data assimilation experiment with $\ensembleSize = 50$ ensemble members for each method. The RMSEs are the caption brackets. 
All three ensemble-based data assimilation methods lead to means that closely resemble the KF reference solution. 
The mean error is in general low and smoothly distributed for both IEWPF and LETKF, whereas the 
errors of EKTF are somewhat bigger.
Based on RMSE, IEWPF performs slightly better than LETKF, whereas the RMSE for ETKF is about twice that of IEWPF.

To deduce reliable conclusions beyond one data set and single ensembles, we repeat the data assimilation experiment multiple times for several independently generated true states. 
In \Cref{tab:advecdiff}, we report averaged results for 20 replicate synthetic truths and five ensemble-based data assimilation experiments each. 
For this relatively small ensemble size of $\ensembleSize=50$, the localisation in the LETKF halves the RMSE compared to the standard ETKF, and the RMSE of the IEWPF lies in the middle of the ETKF with and without localisation.

\paragraph{Frobenius covariance difference}
We contrast the empirical covariance estimates $\widehat{\CovMatrix}^a$ in \cref{eq:ensCovEstimate} with the KF reference $\CovMatrix^a$.
We compute the Frobenius covariance difference (FCD) to compare these covariance matrices:
\begin{equation*}
    \text{FCD} = \| \CovMatrix^a - \widehat{\CovMatrix}^a \|_\text{F},
\end{equation*}
where $\|\cdot\|_\text{F}$ denotes the Frobenius norm (elementwise sum). 

Averaged results for the FCD over replicate experiments are presented in \Cref{tab:advecdiff} using $\ensembleSize=50$.
Here, we see that the FCD for IEWPF and LETKF are very similar for all cases. 
The covariance approximations show smaller errors for the ETKF solution than for the other methods. At a single step, the ETKF approximation to the covariance is unbiased, while the other methods have no such guarantee. Still, it is not obvious that the ETKF performs better after many data assimilation steps. Also, when we test the entries close and far from the diagonal of the covariance matrix, we cannot see any other behaviour in the results.

\paragraph{Integrated quadratic distance}
We next study a metric for the marginal distribution mismatch of discrete ensemble-based distribution approximations to the Gaussian KF reference solution.
The reference cumulative distribution function (CDF) of the KF is denoted $F^a$. The empirical cumulative distribution function (ECDF) of the ensemble-based solutions are denoted $\widehat{F}^a$.

In the analysis two specific locations shown in the far right panel of \Cref{fig:advecdiff-truth} are studied based on their different characteristics: $\loc_1 = (0, 0)$ is an observation site and $\loc_2 = (2.5, 1.5)$ is as far away from observation data as possible.

%
\begin{figure}[h!t]
	\centering
	\begin{subfigure}[t]{0.45\textwidth}
		\begin{tikzpicture}
		\begin{axis}[
		width=0.75\textwidth,
		height=0.3\textwidth,
		scale only axis,
        legend pos=north west,
 		tick pos = left, 
  		xmin=13.75,
  		xmax =14.5,
  		xtick={14, 14.5},
		]
		
		\addplot[color = {rgb:red,170;green,51;blue,119}, line width=0.5pt] table [x index={0}, y index={1}] {files/cdf.txt};
		\addlegendentry{\tiny KF}
		
		\addplot[const plot, dotted, color = {rgb:red,68;green,119;blue,170}, line width=0.5pt] table [x index={3}, y index={0}]  {files/ecdf.txt};
		\addlegendentry{\tiny IEWPF}
		
		\addplot[const plot, dashed, color = {rgb:red,68;green,119;blue,170}, line width=0.5pt] table [x index={1}, y index={0}] {files/ecdf.txt};
		\addlegendentry{\tiny ETKF}

		\addplot[const plot, solid, color = {rgb:red,68;green,119;blue,170}, line width=0.5pt] table [x index={2}, y index={0}] {files/ecdf.txt};
		\addlegendentry{\tiny LETKF}
		
		\end{axis}
		\end{tikzpicture}
		\caption{\quad CDF and ECDFs at $\loc_1$}
	\end{subfigure}%
	\hfill
		\begin{subfigure}[t]{0.45\textwidth}
		\begin{tikzpicture}
		\begin{axis}[
		width=0.75\textwidth,
		height=0.3\textwidth,
		scale only axis,
        legend pos=north west,
 		tick pos = left, 
  		xmin=12.0,
  		xmax =12.75,
  		xtick={12,12.5,13},
		]
		
		\addplot[const plot, dashed, color = {rgb:red,68;green,119;blue,170}, line width=0.5pt] table [x index={4}, y index={0}] {files/ecdf.txt};

		\addplot[const plot, solid, color = {rgb:red,68;green,119;blue,170}, line width=0.5pt] table [x index={5}, y index={0}] {files/ecdf.txt};

		\addplot[const plot, dotted, color = {rgb:red,68;green,119;blue,170}, line width=0.5pt] table [x index={6}, y index={0}]  {files/ecdf.txt};
		

		\addplot[color = {rgb:red,170;green,51;blue,119}, line width=0.5pt] table [x index={0}, y index={2}] {files/cdf.txt};

		\end{axis}
		\end{tikzpicture}
		\caption{\quad CDF and ECDFs at $\loc_2$}
	\end{subfigure}%
	\caption{For two distinct positions, $\loc_1$ as observation site and $\loc_2$ far away from observation sites, at $t=250$ the CDF of the Kalman filter is compared to the ECDFs of IEWPF [0.0242, 0.0238], ETKF [0.0164, 0.0254], and LETKF [0.0093, 0.0117], respectively, with $\ensembleSize=50$. In brackets, the integrated quadratic difference $d_{\text{IQ}}$ for $\loc_1$ and $\loc_2$, respectively.}
	\label{fig:advecdiff-ecdf}
\end{figure}
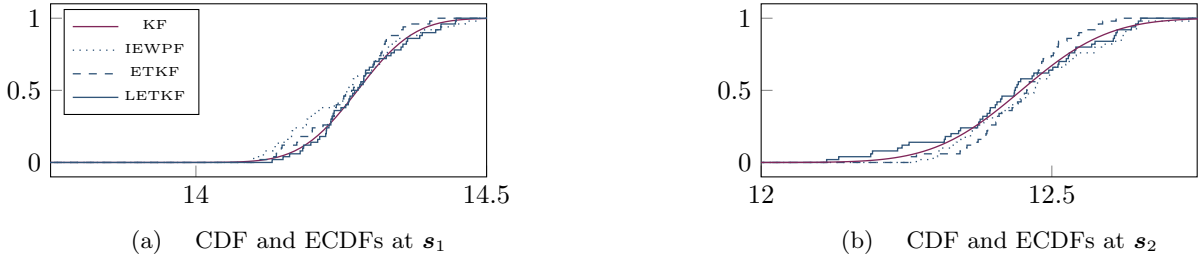

In \Cref{fig:advecdiff-ecdf}, the CDF of the KF is depicted in comparison with the ECDFs of IEWPF, LETKF, and ETKF at the two different locations for a small ensemble size.
First, since the scales of the $x$-axis in both displays are the same, it becomes obvious that the standard deviation at an observation site is much smaller than at an unobserved location. 
Next, we see that the different filtering methods differ in quality when compared to the analytic solution.
For the observation site $\loc_1$, there is no clear qualitative difference, but at $\loc_2$ one may already identify a slight divergence in ETKF's ECDF, whereas LETKF and IEWPF still approximate the reference CDF quite well.

The tuning parameter $\beta$ steers the spread in the analysis ensemble of the IEWPF and among the presented evaluation metrics the ECDF reveals the scale the best. We used it to optimise the filtering distribution manually and found $0.55$ as best choice.
For smaller values, the ECDF gets too sharp and for higher values the spread gets too large.
Similarly, the variance in the LETKF-ensemble usually increases as the inflation parameter $\phi$ decreases. 
When using $\phi<1$ we observed that the spread in the ensemble becomes too big compared to the CDF and the best match is achieved for $\phi=1$. 

\citet{Thorarinsdottir2013} suggest a proper divergence function to compare marginal CDFs, condensing the error into a scalar number.
The integrated quadratic difference is defined by
\begin{equation*}
    d_\text{IQ} = \int{\left(F^a-\widehat{F}^a\right)^2 \mathrm{d}x},
\end{equation*}
where the quadratic error is integrated over the sample space of the variable. Errors captured in $d_\text{IQ}$ can originate from either a lack of Gaussianity, or a wrong scaling or a bias, or a combination. 


\Cref{tab:advecdiff} shows averaged results for $d_{\text{IQ}}$ at $\loc_1$ and $\loc_2$ for the three ensemble methods. IEWPF and ETKF produce similar results, while LETKF clearly gives the best results. The reason is twofold: i) The IEWPF and ETKF update the entire field at each data assimilation time, even at locations like $\loc_2$ that are far away. With the limited ensemble size, this is likely to induce some undesired bias and variability far from data. ii) The LETKF is rather accurate near the observations sites, like the other filters, and because the advection and diffusion are known, the local updating propagates reasonably over time to the far location $\loc_2$.

\paragraph{Probability coverage level}
Based on the mean and variance of the ensemble-based solutions, we can check how often a prediction interval covers the true realisation. For the KF reference, we will have near nominal coverage because the truth is simulated from the same model. 
Coverage probabilities (CPs) of the analysis after the first observation time are
\begin{equation*}
    CP^1_{1.64} \coloneqq P\Big(\state^{1}_{\text{true}} \in \left[ \mean^{1,a} \pm 1.64 \bm{\sigma}^{1,a} \right]\Big) \approx 0.90.
\end{equation*}
This means that the probability that the truth is covered in the interval of 1.64 standard deviations from the mean is 90\%. For all methods, we use replicated synthetic truths, and average the CPs.

\begin{figure}[h!tb]
    \centering
    \hspace*{-0.5cm}
    \begin{subfigure}{0.25\textwidth}
    \includegraphics[]{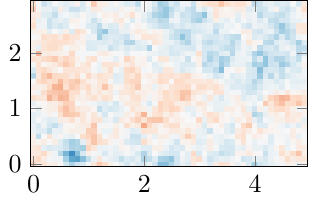}
    \caption{\quad KF $[89.7\%]$}
    \end{subfigure}
    \hfill
    \begin{subfigure}{0.25\textwidth}
    \includegraphics[]{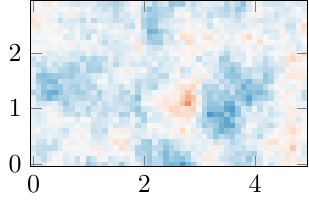}
    \caption{\quad IEWPF $[88.7\%]$}
    \end{subfigure}
    \hfill
    \begin{subfigure}{0.25\textwidth}
    \includegraphics[]{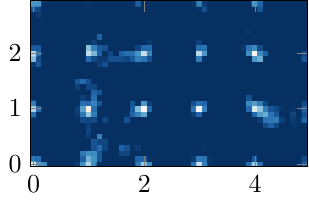}
    \caption{\quad ETKF $[77.8\%]$}
    \end{subfigure}
    \hfill
    \begin{subfigure}{0.25\textwidth}
    \includegraphics[]{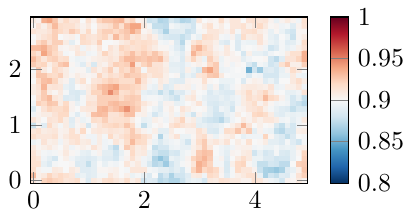}
    \caption{\quad LETKF $[90.4\%]$}
    \end{subfigure}
    \caption{Estimated $CP^1_{1.64}$ using 500 replication experiments for the KF and the ensemble-based methods with $\ensembleSize=50$. 
    The brackets show the averaged estimated coverage probabilities. The color scale is centered at the target probability of 90\%.}
    \label{fig:advecdiff-coverage}
\end{figure}

\Cref{fig:advecdiff-coverage} shows the estimates of the CPs averaged over 500 runs.
The KF results cover the nominal 90\% very well, with a variability as expected from 500 replicates.
The IEWPF and the LETKF also give very good estimates even though the ensemble size of 50 members is low, and there are no visible structures around observation sites.
In contrast, the ETKF without localisation suffers from strong under-coverage in this experiment. 
The CPs are around 0.9 near observation sites but fall to lower levels away from these.

\paragraph{Spatial connectivity}
While the previous metrics have considered the marginal solution at one time step only, the correlation between different time steps and between different spatial positions give further insight into the statistical quality of the filtering methods.

Given data up to time $\simTime\prevDAstep$, the correlation between the concentration at $\loc_k$ at $\simTime^{n-1}=225$ and $\loc_l$ at $\simTime^n$ can then be calculated from the KF results via 
\begin{equation}
    \Corr\left(\state\prevDAstep[a]_k, \state\DAstep[f]_l\right) = \frac{\Cov\left(\state\prevDAstep[a]_k, \state\DAstep[f]_l\right)}{\sigma^{a,n-1}_k \sigma^{f,n}_l} = \frac{\modelMatrix\CovMatrix^{a,n-1}_{k,l}}{\sqrt{\CovMatrix^{a,n-1}_{k,k}} \sqrt{\CovMatrix^{f,n}_{l,l}}}.
\end{equation}
Similarly, we can estimate these correlations from the ensemble-based methods by
\begin{equation*}
    \widehat{\Corr}\left(\state\prevDAstep[a]_k, \state\DAstep[f]_l\right) =
    \frac{1}{\ensembleSize-1} \frac{1}{\hat{\sigma}\prevDAstep[a]_k \hat{\sigma}\DAstep[f]_l} \sum_{e=1}^{\ensembleSize}{ ({\state\prevDAstep[a]_{e,k}} - \overline{\state}\prevDAstep[a]_{k}) ({\state\DAstep[f]_{e,l}} - \overline{\state}\DAstep[f]_{l}) }.
\end{equation*}

\begin{figure}
    \centering
    \hspace*{-0.5cm}
    \begin{subfigure}[b]{0.25\textwidth}
        \includegraphics[]{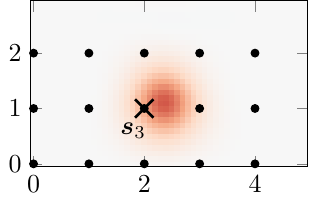}
        \includegraphics[]{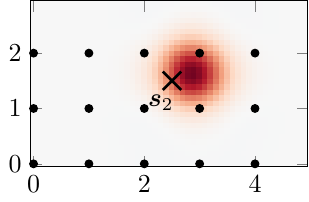}
        \caption{\quad KF }
    \end{subfigure}
    \hfill
    \begin{subfigure}[b]{0.25\textwidth}
        \includegraphics[]{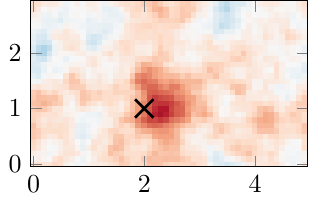}
        \includegraphics[]{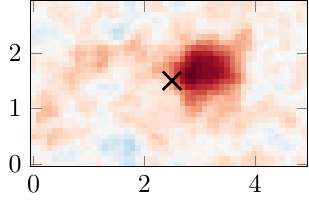}
        \caption{\quad IEWPF$[2.73, 2.71]$}
    \end{subfigure}
    \hfill
    \begin{subfigure}[b]{0.25\textwidth}
        \includegraphics[]{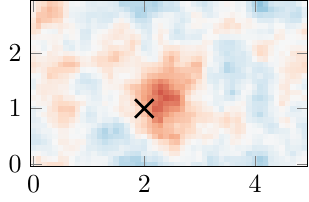}
        \includegraphics[]{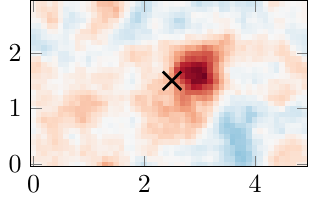}
        \caption{\quad ETKF $[2.32, 2.64]$}
    \end{subfigure}
    \hfill
    \begin{subfigure}[b]{0.25\textwidth}
        \includegraphics[]{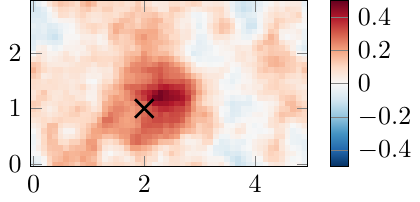}
        \includegraphics[]{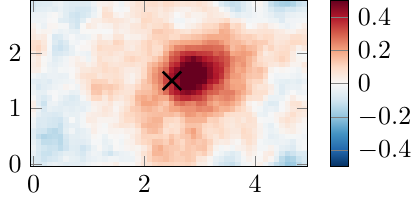}
        \caption{\quad LETKF $[3.86, 3.13]$}
    \end{subfigure}
    \caption{Correlations $\Corr\left(\state^9_k,\state^{10}_l\right)$ and $\widehat{\Corr}\left(\state^9_k,\state^{10}_l\right)$ between fixed locations $\loc_k$ and all other grid points in the domain $\loc_l$, for $k = 3 $ in the upper and $k=2$ in the lower row, respectively. The ensemble-based methods use $\ensembleSize = 250$ ensemble members, and the respective CE estimates are given in the square brackets.}
    \label{fig:advecdiff-correlations}
\end{figure}

In \Cref{fig:advecdiff-correlations}, we show the correlation fields of the state between a reference grid cell at $t^{9}=225$ and all other grid cells at $t^{10}=250$. 
As reference locations $\loc_k$, we consider $\loc_2$ positioned far away from any observations in the top row, and $\loc_3$ at an observation site in the middle of the domain displayed in the lower row.
First, from the KF solution (left), we recognise the advection field in the model that transports information towards east-north-east from both locations, as well as the diffusion causing the correlation to have longer range than the model error correlation radius.
Second, the maximal correlation to the reference point is higher when $\loc_k$ is not an observation site. 
In spatial statistics, conditioning on data breaks up some of the prior correlations. 
Since most of the update from the data assimilation occurs near the observation locations, the conditional correlation tends to be smaller in the proximity of data. 
At locations that are far from observations, more of the prior correlation remains. 

In the three rightmost columns of \Cref{fig:advecdiff-correlations} we see the correlations estimated with the three ensemble-based methods using an ensemble with $\ensembleSize=250$ members.
We see that all methods capture a similar correlation structure with respect to the advection and diffusion, and the relative balance between prior model and information from the observation. With $\ensembleSize=250$, results are less smooth than the KF solution.
In the upper scenario the area of high analytical correlations becomes less apparent among background noise, while in the lower scenario the respective regions are easier to identify in all methods. 

The spatial error in the approximation of correlation between two consecutive model steps for the reference location $\loc_k$ is evaluated collectively across all grid cells as 
\begin{equation}
    \text{CE}(\loc_k)^2 = \sum_{l=1}^{\gridSize}{ \left| \Corr\left(\state\prevDAstep[a]_k, \state\DAstep[f]_l\right) - \widehat{\Corr}\left(\state\prevDAstep[a]_k, \state\DAstep[f]_l\right) \right|^2 }.
    \label{eq:CE}
\end{equation}
The CE for the specific data assimilation run shown in \Cref{fig:advecdiff-correlations} are given in square brackets in the figure captions. 
These results for a single run with $\ensembleSize=250$ already reveal that the contribution to CE can come from multiple sources, such as over- or under-estimation of the actual correlations and from spurious correlations. 
The final CE does not qualitatively expose which of the error sources are present to which extent, but it quantifies how well the analytical structure is approximated.
The chosen $\Corr$ and $\widehat{\Corr}$ suggest that spurious correlations are a present error source in all methods, but IEWPF and LETKF tend to overestimate the high correlations. Meanwhile, ETKF underestimates the correlations, thus leading to a smaller CE.


\subsection{Discussion of evaluation metrics}

The set of comparative metrics from the previous subsection has given us a collection of metrics that quantify some statistical qualities of the ensemble-based data assimilation methods in reference to the analytical KF solution.



\begin{table}
    \centering
    \begin{tabular}{ p{2cm}||p{2cm}||p{2cm}||p{1.5cm}|p{1.5cm} }
        & RMSE & FCD & $d_{IQ}$ at $\loc_1$ & $d_{IQ}$ at $\loc_2$ \\
        \hline
        MC    & 8.27 [2.88] &  47.0 [8.35] & 12.8E-02 & 13.5E-02\\
        IEWPF & 1.67 [0.43] &  2.77 [0.14] & 2.51E-02 & 2.58E-02 \\
        ETKF  & 2.14 [0.40] &  2.14 [0.04] & 2.57E-02 & 2.86E-02 \\
        LETKF & 1.15 [0.24] &  2.79 [0.15] & 1.29E-02 & 1.68E-02 \\
         \hline
        \end{tabular}
    \caption{Metrics for marginal distribution averaged over 20 synthetic truths and 5 ensemble idealisations each with $\ensembleSize=50$. In brackets the standard deviations.}
    \label{tab:advecdiff}
\end{table}

\Cref{tab:advecdiff} shows the statistically averaged results for these performance scores at $\simTime=250$. These results are obtained across five data assimilation runs for 20 different synthetic truths and are therefore more reliable than the single realisations demonstrated in \Cref{fig:advecdiff-coverage,fig:advecdiff-rmse,fig:advecdiff-ecdf}.
We have here used $\ensembleSize = 50$ ensemble members for each run.
In addition to comparing the data assimilation methods against each other, we have also included the results using pure Monte Carlo simulations without observations (top row). These serve to demonstrate the worst-case scenario for each metric, and we see how all three data assimilation methods clearly outperform this, as expected. 
In the experiments, we have observed that the IEWPF takes several assimilation steps until it is sufficiently calibrated, what is respected by the the choice of $t$ here such that the comparison stays fair, see \Cref{sec:gpuocean} for details.


Based on the results in \Cref{tab:advecdiff} there is no method that clearly dominate on all individual criteria. 
For RMSE, it seems that LETKF is much better than ETKF, but this is not as clear when considering FCD, where ETKF scores best.
Maybe more surprising, LETKF is significantly better than ETKF when measuring the error in the ECDF at $\loc_2$ far from the observation, but not at $\loc_1$ at an observation site. A plausible explanation is that when updating the state far from an observation, all covariances are relatively weak, which means that spurious correlations more easily dominate data assimilation. 
With localisation, we ensure that only the most relevant small correlations are considered, thus improving the result.
This effect will then be less at an observation site as the most important correlations are stronger.
IEWPF gets all metrics between ETKF and LETKF.
We observe that worse FCD has no influence on the $d_{\text{IQ}}$ at the considered positions.

\paragraph{Effects of ensemble size}
\begin{figure}[h!t]
	\centering
	\begin{subfigure}[t]{0.3\textwidth}
	\centering
		\begin{tikzpicture}
		\begin{axis}[
		width=\textwidth,
		height=0.5\textwidth,
		scale only axis,
        legend style={at={(0.5,0.975)},anchor=north},
 		xlabel={$N_e$},
 		xmode = log,
 		tick pos = left, 
 		ymin=0,
 		ymax = 4.0,
		]
		
		\addplot[dotted,mark=diamond*, color = {rgb:red,68;green,119;blue,170}, line width=0.5pt] table [x index={0}, y = avg_mean_rmse_iewpfs,ignore chars={\#}] {files/results_Ne_avg.txt};
		
		\addplot[dashed, mark=diamond*, color = {rgb:red,68;green,119;blue,170}, line width=0.5pt] table [x index={0}, y = avg_mean_rmse_etkfs,ignore chars={\#}] {files/results_Ne_avg.txt};
		
		\addplot[solid,mark=diamond*, color = {rgb:red,68;green,119;blue,170}, line width=0.5pt] table [x index={0}, y = avg_mean_rmse_letkfs,ignore chars={\#}] {files/results_Ne_avg.txt};

		
		

		\end{axis}
		\end{tikzpicture}
		\caption{\quad $\text{RMSE}$ }
	\end{subfigure}%
 	\hfill
	\begin{subfigure}[t]{0.3\textwidth}
	\hspace*{-0.5cm}
		\begin{tikzpicture}
		\begin{axis}[
		width=\textwidth,
		height=0.5\textwidth,
		scale only axis,
 		xlabel={$N_e$},
 		xmode = log,
 		tick pos=left,
 		ymin=0,
 		ymax = 6e-2,
 		legend style={at={(0.5,1.25)},anchor=south},
 		legend columns=3,
		]
		
		\addplot[dotted,mark=diamond*, color = {rgb:red,68;green,119;blue,170}, line width=0.5pt] table [x index={0}, y = ecdf_err_iewpfs0, ignore chars={\#}] {files/results_Ne_avg.txt};
		\addlegendentry{IEWPF~~~~}
		
 		\addplot[dashed,mark=diamond*, color = {rgb:red,68;green,119;blue,170}, line width=0.5pt] table [x index={0}, y = ecdf_err_etkfs0, ignore chars={\#}] {files/results_Ne_avg.txt};
 		\addlegendentry{ETKF~~~~}
		
 		\addplot[solid,mark=diamond*, color = {rgb:red,68;green,119;blue,170}, line width=0.5pt] table [x index={0}, y = ecdf_err_letkfs0, ignore chars={\#}] {files/results_Ne_avg.txt};
 		\addlegendentry{LETKF~~~~}

 		\addplot[dotted,mark=diamond*, color = {rgb:red,170;green,51;blue,119}, line width=0.5pt] table [x index={0}, y = ecdf_err_iewpfs2, ignore chars={\#}] {files/results_Ne_avg.txt};
 		
 		\addplot[dashed,mark=diamond*, color = {rgb:red,170;green,51;blue,119}, line width=0.5pt] table [x index={0}, y = ecdf_err_etkfs2, ignore chars={\#}] {files/results_Ne_avg.txt};
 		
 		\addplot[solid,mark=diamond*, color = {rgb:red,170;green,51;blue,119}, line width=0.5pt] table [x index={0}, y = ecdf_err_letkfs2, ignore chars={\#}] {files/results_Ne_avg.txt};
				
		\end{axis}
		\end{tikzpicture}
		\caption{\quad $d_\text{IQ}$ at $\loc_0$ (blue) and $\loc_2$ (purple)}
	\end{subfigure}%
	 \hfill
	\begin{subfigure}[t]{0.3\textwidth}
	\centering
		\begin{tikzpicture}
		\begin{axis}[
		width=\textwidth,
		height=0.5\textwidth,
		scale only axis,
        legend pos=north east,
 		xlabel={$N_e$},
 		xmode = log,
 		tick pos=left,
 		ymin=0,
 		ymax = 9,
		]
		
		\addplot[dotted,mark=diamond*, color = {rgb:red,68;green,119;blue,170}, line width=0.5pt] table [x index={0}, y index={32}] {files/ensemblesize.txt};
		
		\addplot[dashed,mark=diamond*, color = {rgb:red,68;green,119;blue,170}, line width=0.5pt] table [x index={0}, y index={30}] {files/ensemblesize.txt};
		
		\addplot[solid,mark=diamond*, color = {rgb:red,68;green,119;blue,170}, line width=0.5pt] table [x index={0}, y index={31}] {files/ensemblesize.txt};

		\addplot[dotted,mark=diamond*, color = {rgb:red,170;green,51;blue,119}, line width=0.5pt] table [x index={0}, y index={38}] {files/ensemblesize.txt};
		
		\addplot[dashed,mark=diamond*, color = {rgb:red,170;green,51;blue,119}, line width=0.5pt] table [x index={0}, y index={36}] {files/ensemblesize.txt};
		
		\addplot[solid,mark=diamond*, color = {rgb:red,170;green,51;blue,119}, line width=0.5pt] table [x index={0}, y index={37}] {files/ensemblesize.txt};
		
		\end{axis}
		\end{tikzpicture}
		\caption{\quad $\text{CE}(\loc_3)$ (blue) and $\text{CE}(\loc_2)$ (purple)}
		\label{fig:advecdiff-ensemblesize-ce}
	\end{subfigure}
	\caption{Evolution of the comparison measures in depends of the ensemble size $\ensembleSize$.}
	\label{fig:advecdiff-ensemblesize}
\end{figure}
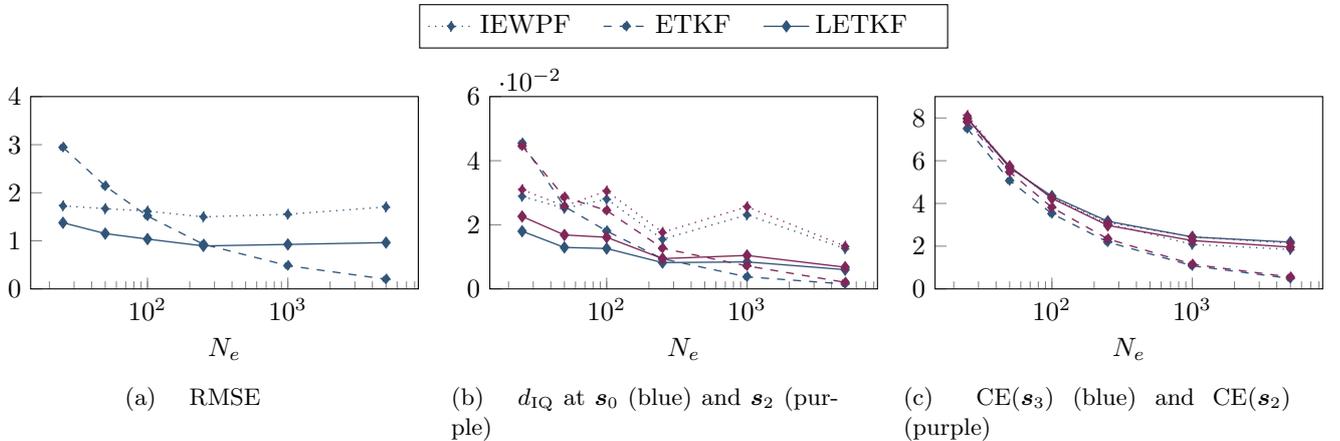

In \Cref{fig:advecdiff-ensemblesize}, we study how different ensemble sizes influence RMSE, $d_{\text{IQ}}$, and correlations for the three ensemble-based methods.
The results are averaged across multiple experiments for each of the ensemble sizes $\ensembleSize\in\{25,50,100,250,1000,5000\}$.
%
%
In general, we expect that increased ensemble size leads to more accurate statistical estimates and thereby better results.
This is clearly the case for ETKF for all metrics, and LETKF shows the same trend, but not as strongly. We see that LETKF outperforms ETKF with respect to RMSE and $d_{\text{IQ}}$ for small ensemble sizes, but ETKF is better with large ensembles as the performance of LETKF stagnates for $\ensembleSize > 250$.
LETKF improves less than ETKF with larger ensembles because it ignores correlations, and this gives bias in the analysis.
The IEWPF yields results between ETKF and LETKF for small sample sizes, but there is slower convergence as the ensemble size increases. Unlike ETKF, which converges to the true Gaussian distribution in this case, there is no such guarantee for the IEWPF. 
Since the second stage perturbation step of the IEWPF is designed to reduce a systematic bias and help performance, fine-tuning the choice of $\beta$ scaling parameter could improve convergence for some properties, but maybe not similarly so for all the desired scores.
The correlations mismatch compared with the KF in \Cref{fig:advecdiff-correlations} are slightly different depending on the fixed reference point, but they converge with increasing sample, especially so for the ETKF which has curves going faster to $0$. For both LETKF and IEWPF there seem to be a remaining mismatch in this CE score even for thousands of ensemble members.

\paragraph{Effects of sparsity of observational data}
In a regime dominated by the sparsity of observations, we also want to stress-test all methods with respect to the amount of observational data.
For this purpose, we repeat the case study 
using $\obsSize\in\{8,15,60,104,170\}$ regularly placed observation sites. These numbers are chosen such that the observation locations have distance of 15, 10, 5, 4, and 3 grid cells apart from each others, respectively. We use $\ensembleSize = 50$ ensemble members.
Of course, the localisation scheme for the LETKF is not designed for dense data and will get computationally very inefficient due to a high number of batches that are processed serially. 
The localisation radius is not modified.

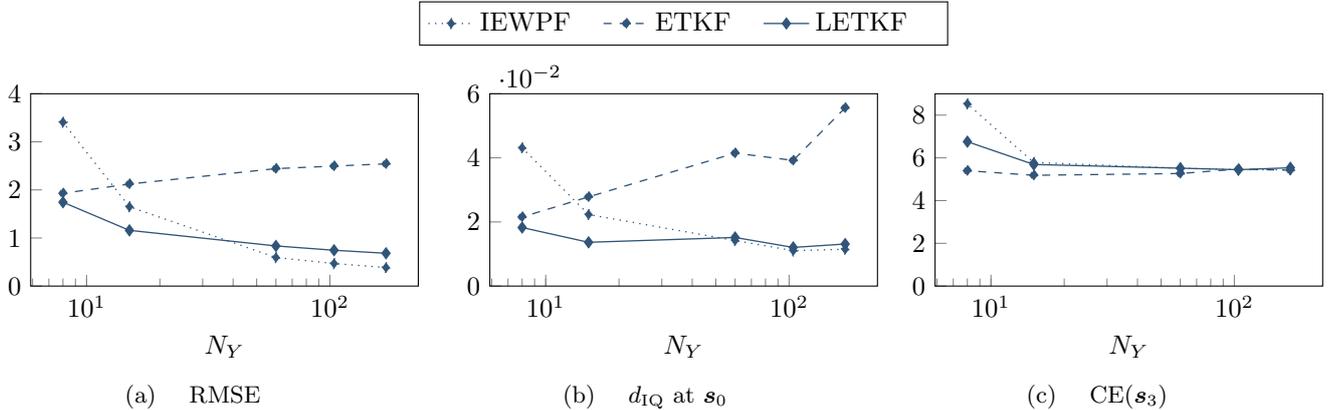
\begin{figure}[h!t]
	\centering
	\begin{subfigure}[t]{0.3\textwidth}
	\centering
		\begin{tikzpicture}
		\begin{axis}[
		width=\textwidth,
		height=0.5\textwidth,
		scale only axis,
        legend style={at={(0.5,0.975)},anchor=north},
        legend cell align={left},
 		xlabel={$\obsSize$},
 		xmode = log,
 		tick pos=left,
 		ymin=0,
 		ymax = 4,
		]
		
		\addplot[dotted, mark=diamond*, color = {rgb:red,68;green,119;blue,170}, line width=0.5pt]  table [x index={0}, y = avg_mean_rmse_iewpfs, ignore chars={\#}] {files/results_Ny_avg.txt};
		
		\addplot[dashed, mark=diamond*, color = {rgb:red,68;green,119;blue,170}, line width=0.5pt] table [x index={0}, y = avg_mean_rmse_etkfs, ignore chars={\#}] {files/results_Ny_avg.txt};
		
		\addplot[solid, mark=diamond*, color = {rgb:red,68;green,119;blue,170}, line width=0.5pt] table [x index={0}, y = avg_mean_rmse_letkfs, ignore chars={\#}] {files/results_Ny_avg.txt};
		
		\end{axis}
		\end{tikzpicture}
		\caption{\quad RMSE}
	\end{subfigure}%
 	\hfill
	\begin{subfigure}[t]{0.3\textwidth}
	\hspace*{-0.5cm}
		\begin{tikzpicture}
		\begin{axis}[
		width=\textwidth,
		height=0.5\textwidth,
		scale only axis,
 		xlabel={$\obsSize$},
 		xmode = log,
 		tick pos=left,
 		ymin=0,
 		ymax = 0.06,
 		legend style={at={(0.5,1.25)},anchor=south},
 		legend columns=3,
		]
		
 		\addplot[dotted,mark=diamond*, color = {rgb:red,68;green,119;blue,170}, line width=0.5pt] table [x index={0}, y = ecdf_err_iewpfs0, ignore chars={\#}] {files/results_Ny_avg.txt};
 		\addlegendentry{IEWPF~~~~}
		
 		\addplot[dashed,mark=diamond*, color = {rgb:red,68;green,119;blue,170}, line width=0.5pt] table [x index={0}, y = ecdf_err_etkfs0, ignore chars={\#}] {files/results_Ny_avg.txt};
 		\addlegendentry{ETKF~~~~}
		
 		\addplot[solid,mark=diamond*, color = {rgb:red,68;green,119;blue,170}, line width=0.5pt] table [x index={0}, y =ecdf_err_letkfs0, ignore chars={\#}] {files/results_Ny_avg.txt};
 		\addlegendentry{LETKF~~~~}

		\end{axis}
		\end{tikzpicture}
		\caption{\quad $d_\text{IQ}$ at $\loc_0$}
	\end{subfigure}%
	 \hfill
	\begin{subfigure}[t]{0.3\textwidth}
	\centering
		\begin{tikzpicture}
		\begin{axis}[
		width=\textwidth,
		height=0.5\textwidth,
		scale only axis,
        legend pos=north east,
 		xlabel={$\obsSize$},
 		xmode = log,
 		tick pos=left,
 		ymin=0,
 		ymax = 9,
		]
		
    	\addplot[dotted,mark=diamond*, color = {rgb:red,68;green,119;blue,170}, line width=0.5pt] table [x index={0}, y index={32}] {files/observationsize.txt};
		
		\addplot[dashed,mark=diamond*, color = {rgb:red,68;green,119;blue,170}, line width=0.5pt] table [x index={0}, y index={30}] {files/observationsize.txt};
		
		\addplot[solid,mark=diamond*, color = {rgb:red,68;green,119;blue,170}, line width=0.5pt] table [x index={0}, y index={31}] {files/observationsize.txt};

		\end{axis}
		\end{tikzpicture}
		\caption{\quad $\text{CE}(\loc_3)$}
	\end{subfigure}
	\caption{Evolution of the comparison measures in dependence of the sparsity/density of observation data $\obsSize$. Fixed ensemble size $\ensembleSize=50$.}
	\label{fig:advecdiff-obssize}
\end{figure}

Figure \Cref{fig:advecdiff-obssize} shows the same averaged metrics as before, with respect to a growing number of observations. Note that it no longer makes sense to distinguish between locations near and far from observations, since the observation sites get denser over the domain. 
The increase in observation data leads to a sharpening in the reference distribution calculated from the KF.
For the ETKF,  we observe that RMSE does not improve and its ECDF approximation gets worse, compared with the KF. This is because of the underestimation in variance and a slight bias which strongly penalises the $d_{\text{IQ}}$. 
Both LETKF and IEWPF improve their quality for increasing observation data size. 
This is surprising and noteworthy for IEWPF, as PFs tends to collapse for high dimensional observations.
In contrast to RMSE and $d_{\text{IQ}}$, the CE does not depend on the observation sparsity and is practically constant on the level that we saw in \Cref{fig:advecdiff-ensemblesize-ce}.

\paragraph{Summary}
In this case study, we consider a linear Gaussian data assimilation problem which gives us the opportunity to verify the ensemble-based methods from \Cref{sec:DA} against the analytical KF reference.
Concluding from these results, we first report that the presented localisation scheme for sparse data presented in \Cref{sec:DA} works properly.
Beyond the verification, we can in particular record that LETKF outperforms ETKF and IEWPF for smaller ensemble sizes (about $\ensembleSize\leq250)$. 
While the ETKF requires a large ensemble size to obtain a reasonable approximation of the full covariance matrix,
the localised version that ignores large-distance correlations is performing well for small ensemble sizes, and it does not improve much more for larger sizes. Similar tendencies are seen with the IEWPF.
The approximation of the correlations between different time steps depend mostly on the ensemble size - the model error plays a major role in the evaluation and this criterion requires a higher ensemble size for a sufficient representation. 
For reasonable ensemble sizes, say $100$, both IEWPF and LETKF operate well for any density of observation data. 
For most criteria we tested in this example with sparse point data, the LETKF tends to give slightly better performance than the IEWPF.
Based on this extensive statistical evaluation, we hence recommend considering the LETKF for similar kinds of applications with sparse data and limited ensemble sizes.

\section{Comparison for Drift Trajectory Forecasting in a Simplified Ocean Model}
\label{sec:gpuocean}


We now increase both dimensionality and complexity as we turn to a non-linear simplified ocean model.
This gives insight into the behaviour of the ensemble-based data assimilation methods on a challenging case with applied relevance.
The practical purpose of this configuration is to use ensembles of computationally efficient simplified ocean models instead of or complementary to single realisations of complex operational ocean models in time critical situations. The simplified models allow for larger ensembles and hence facilitate uncertainty quantification.
Such an approach can be useful in search-and-rescue operations, where drifters released by the vessel or relevant anchored buoys (also called moorings) can give sparse in-situ observations during the operation. 
These point observations can then be assimilated into the ensemble-based representation to improve the drift trajectory forecasts that specify a search area. 

Due to the non-linearity of such an model, there is no analytical reference solution for the ensemble distributions available. We can nevertheless compare LETKF and IEWPF by studying their predictive properties with the ground truth in a simulation study. 
We base our numerical experiments on those presented in \citet{Holm2020c}, where the IEWPF was successfully tailored for efficient GPU-accelerated assimilation of point observations of a chaotic shallow-water model. 
We expand on the numerical result from that work by evaluating more skill scores, and by providing an in-depth comparison between IEWPF and LETKF for state estimation and drift trajectory forecasts.
This will not only serve to assess our novel localisation scheme in the LETKF, but also give a more thorough evaluation of the applicability of IEWPF in this context.




\subsection{Simplified ocean model}

The simplified ocean model is represented by the rotational shallow-water equations given by
\begin{equation}
    \begin{aligned}
        \eta_t + (hu)_x + (hv)_y &= 0 \\
        (hu)_t + \left( hu^2 + \frac{1}{2} gh^2 \right)_x + (huv)_y &= fhv \\
        (hu)_t + (huv)_x + \left( hv^2 + \frac{1}{2}gh^2 \right)_y &= -fhv.
    \end{aligned}
    \label{eq:swe}
\end{equation}
This is a non-linear two-dimensional hyperbolic conservation law, which models conservation of mass through the deviation $\eta$ from equilibrium sea level, and conservation of momentum through $hu$ and $hv$, which are vertically integrated ocean currents in $x$- and $y$-direction, respectively. By denoting the equilibrium depth of the ocean by $H$, we get the total depth as $h = H + \eta$.
Furthermore, $g$ is the gravitational constant and $f$ is the Coriolis parameter that accounts for the rotating frame of reference.

We solve \cref{eq:swe} using the high-resolution central-upwind finite-volume scheme proposed by \citet{Chertock2018}. 
In our notation from the model equation \cref{eq:DAmodel}, the state vector $\state^n$ consists of the cell averaged values $(\eta_i^n, (hu)_i^n, (hv)_i^n)$ at time $\simTime\DAstep$ for all cells $i$ in the discretised domain. The $\modelOperator\DAstep$ operator then applies the finite-volume scheme to evolve the state from $\state\prevDAstep$ to $\state\DAstep$.
Note that the time step used by the numerical method can be chosen independently from the model time step, meaning that $\modelOperator$ might consist of multiple iterations of the numerical scheme.

We apply a small-scale Gaussian model error $\modelError \sim \mathcal{N}(0,\modelErrorCov)$.
It is constructed from a coarse-scale perturbation of $\eta$, which is smoothed by a second-order autoregressive function and projected onto the numerical grid.
The model error for $hu$ and $hv$ is then inferred according to geostrophic balance to ensure physical feasibility.
Further details about this model are available in \citet{Brodtkorb2021} and \citet{Holm2020c}.



\subsection{Experiment design}

In the following, we use the same experimental design as in \citet{Holm2020c}.
We consider a rectangular domain covering $\SI{1100}{\kilo\meter}\times\SI{666}{\kilo\meter}$ that is discretised as a uniform Cartesian $500\times300$ grid. 
The domain has periodic boundary conditions and constant equilibrium depth $H=\SI{230.0}{\meter}$.
The initial conditions, for the ground truth as well as for all ensemble members, consist of a westward jet in the north part of the domain and an eastward jet in the south, with $hv = 0$. Both jets are balanced according to geostrophy by $\eta$ so that the initial conditions are in steady state.
This steady state is however unstable, and slight perturbations, such as those from the model error $\modelError$, cause chaotic behaviour. 

\begin{figure}
    \centering
    \hspace*{-1cm}
    \begin{subfigure}[b]{0.2\textwidth}
        \includegraphics[height=3cm, trim=0cm 0cm 1.5cm 1cm, clip]{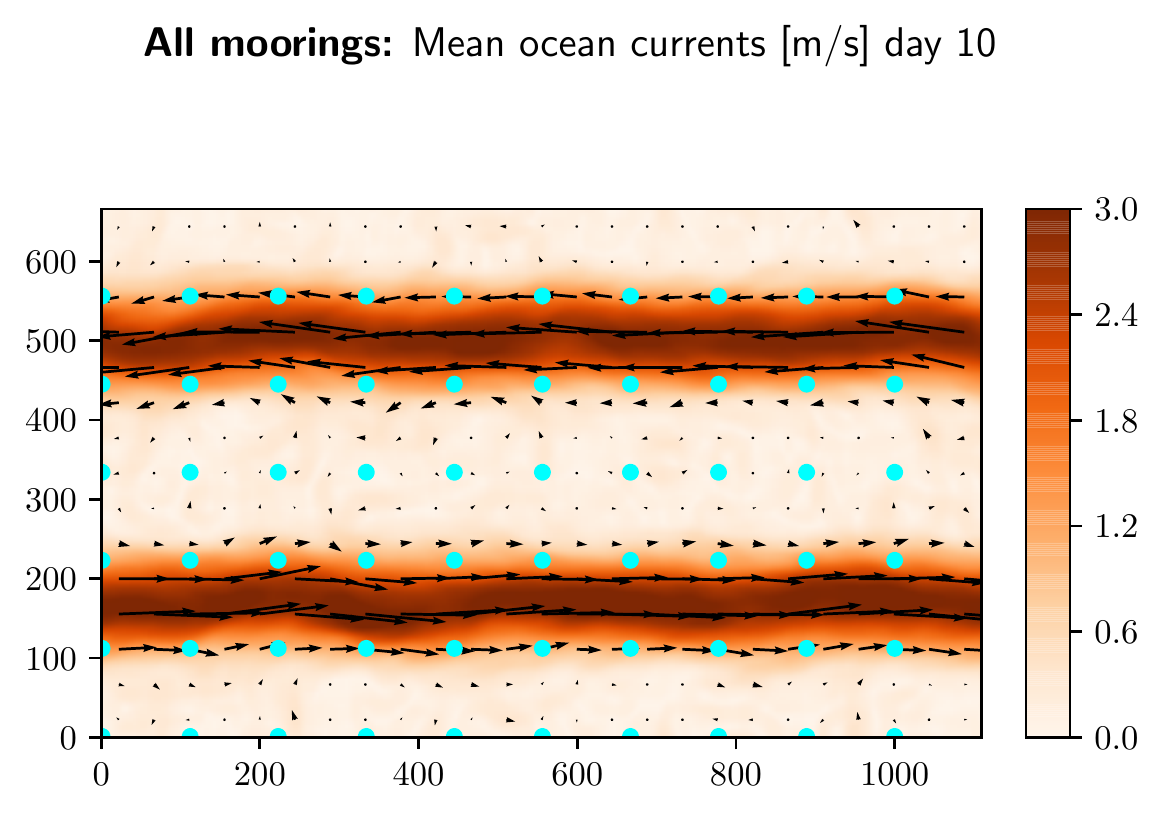}
        \caption{\quad Day 3}
    \end{subfigure}
    \hspace{1.5cm}
    \begin{subfigure}[b]{0.2\textwidth}
        \includegraphics[height=3cm, trim=0cm 0cm 1.5cm 1cm, clip]{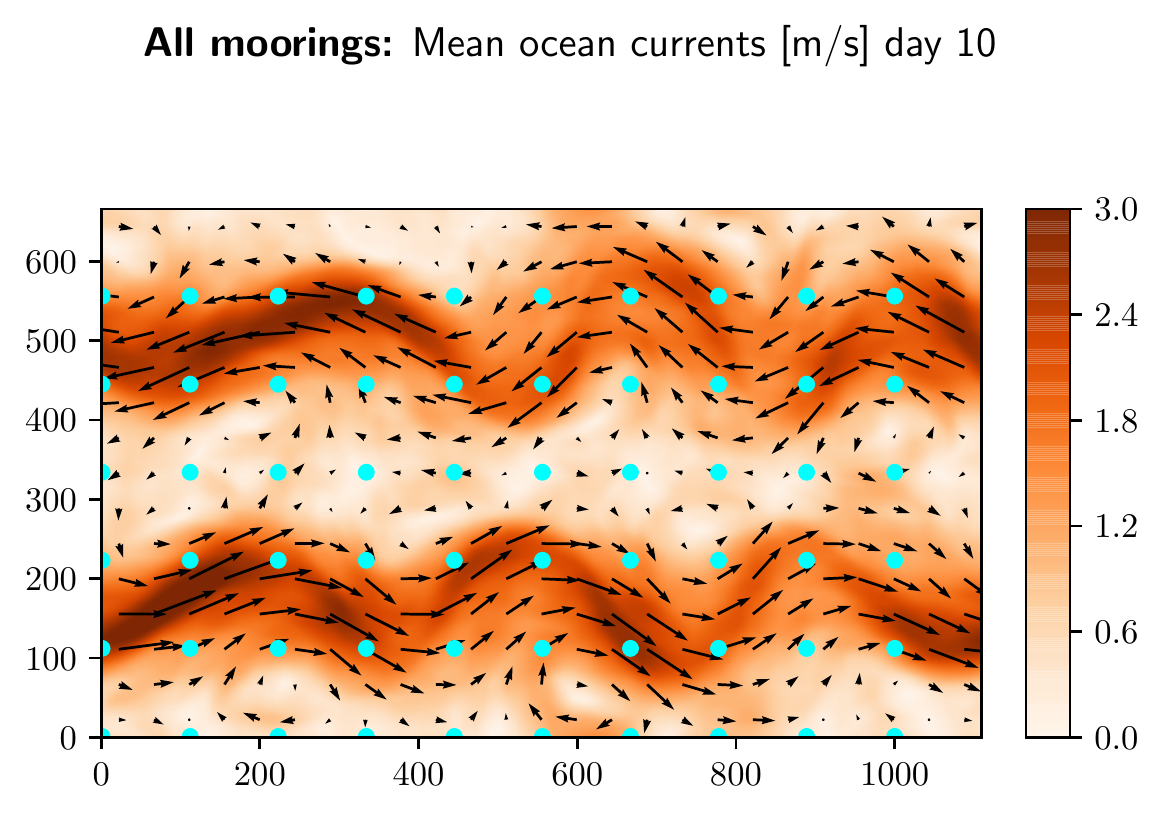}
        \caption{\quad Day 6}
    \end{subfigure}
    \hspace{1.5cm}
    \begin{subfigure}[b]{0.2\textwidth}
        \includegraphics[height=3cm, trim=0cm 0cm 0cm 1cm, clip]{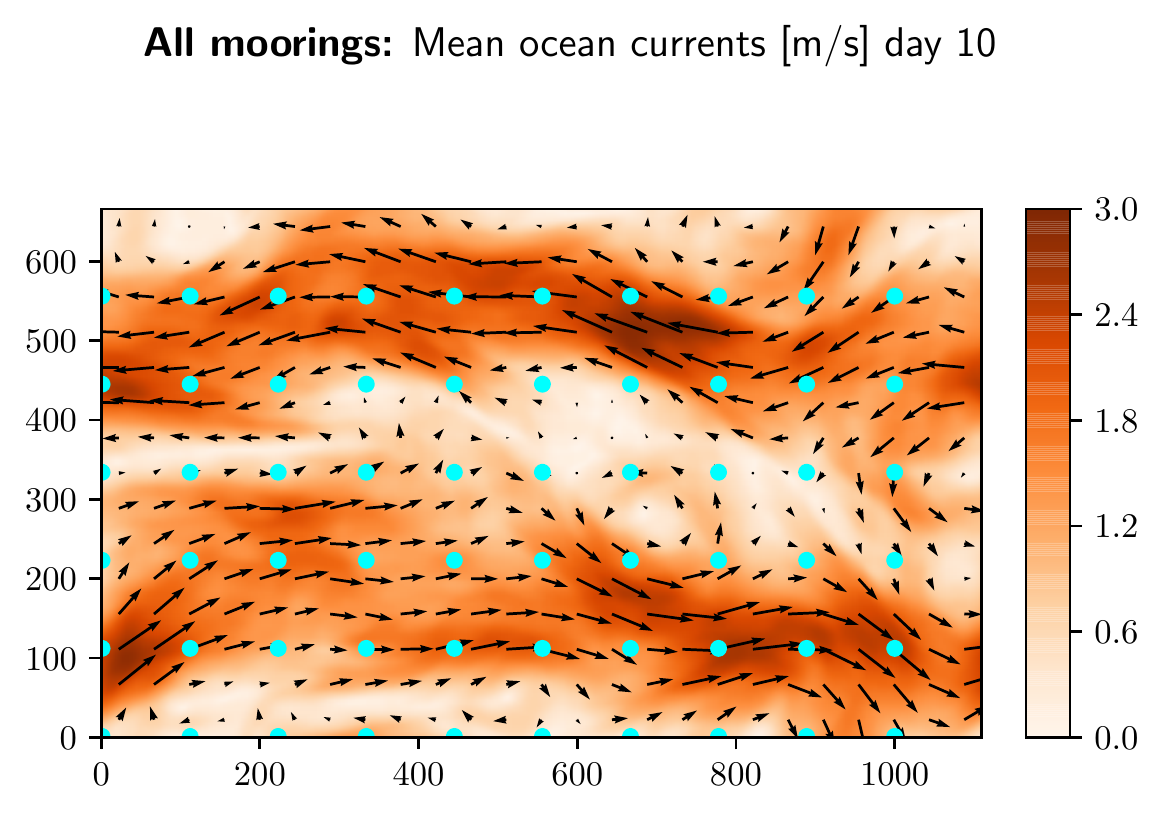}
        \caption{\quad Day 10}
    \end{subfigure}
    \caption{State of the synthetic truth after 3, 6, and 10 simulation days. The arrows indicate the direction as well as strength of the ocean currents derived from $u$ and $v$, respectively. The background visualises the magnitude of velocity in $\SI{}{\meter\per\second}$. The turquoise dots mark the fixed-point buoy positions.}
    \label{fig:gpuocean-truth}
\end{figure}

As an example of the turbulent behaviour, \Cref{fig:gpuocean-truth} shows the water velocities for one realisation that is labeled as the synthetic truth $\state_\text{true}$.
Here, the model error is added every $\SI{60}{\second}$, and the model error correlation radius is approximately 40~km.
From \Cref{fig:gpuocean-truth}, we see that the jets in $\state_\text{true}$ are still quite regular after 3 days, but grow more irregular after 6 and 10 days. It should be noted that the mean state from a pure Monte Carlo experiment without data assimilation will results in $hv \approx 0$ even after 10 days.
This indicates that it is challenging to correctly capture where and how the turbulent behaviour will develop.

From $\state_\text{true}$, we extract direct observations of only $(hu, hv)$ at 60 locations in the domain every 5~minutes between day 3 and day 10, with observational noise sampled from $N(0, \mathbf{I})$.
The turquoise dots in \Cref{fig:gpuocean-truth} show the observation sites.
In total, the experiment is characterised by 450.000 state variables versus only 120 very sparse noisy observations.
After day 10, three~drifters are released in the domain, and advected according to the simulated currents at every time step of the numerical scheme using a simple Euler scheme.
Part of the challenge for the data assimilation methods is to forecast the trajectories of these drifters.

The data assimilation start at simulation day 3 after each ensemble member has been spun up from the steady state through independently sampled model errors.
Even though all ensemble members are visually very similar at this stage, they have started to develop internal instabilities that will grow over time unless the observations are successfully assimilated.



This case is much more challenging than the advection-diffusion model in \Cref{sec:advectiondiffusion}:
The shallow-water model is non-linear, there are unobserved variables, and it has significantly higher dimensionality.
Critically, the non-linear dynamics of the shallow-water model is challenging to capture. In the advection-diffusion model, the state converges towards an equilibrium due to the diffusion, whereas our shallow-water equation case gets chaotic dynamics that makes the ensemble naturally diverge in time.

\subsection{Numerical results}
\label{sec:gpuoceanNumResults}



Classical EnKF approaches like the ETKF lead to useless results for this difficult case, and only results of the IEWPF and LETKF are shown in the comparison. We use $\ensembleSize=100$ as a compromise between computational effort and statistical quality.
Based on our experiments on this high-dimensional non-linear model, the IEWPF performance is not very sensitive to the explicit choice of $\beta$ and we use the maximal allowed value.
In contrast to \Cref{sec:advectiondiffusion}, we now also investigate the influence of inflation in the LETKF.
We present results for the LETKF without inflation ($\phi=1.0$) and for the LETKF whose weights in the localisation are scaled by $\phi=0.5$. 

We compare data assimilation methods with the simulated truth. In this comparison, we use a number of skill scores that refer to the ground truth.

\paragraph{State estimation}
We first look at deviations of the ensemble mean from the truth by
\begin{equation}
    \textbf{err}_\text{mean}^{\text{day10, true}} = \overline{\state}^{\text{day10}} - \state_\text{true}^{\text{day10}},
\end{equation} 
which represents the error in the correct physical unit.
We also investigate the standard deviation in the ensemble 
\begin{equation}
    \text{STD}^\text{day10} = \frac1{\ensembleSize-1} \sqrt{ \sum_{e = 1}^{\ensembleSize}{\left( \state^{\text{day10}}\member - \overline{\state}^{\text{day10}} \right)^2} }
\end{equation}
which 
gives insight about the ensemble spread around its mean.

\begin{figure}
    \centering
    \begin{subfigure}{0.8\textwidth}
        \centering
        \rotatebox{90}{\small {\parbox{2cm}{(i) IEWPF \\ \textcolor{white}{()}}} } \quad
        \includegraphics[width=0.8\textwidth, trim=0.9cm 0.6cm 0cm 1.5cm, clip, valign=b]{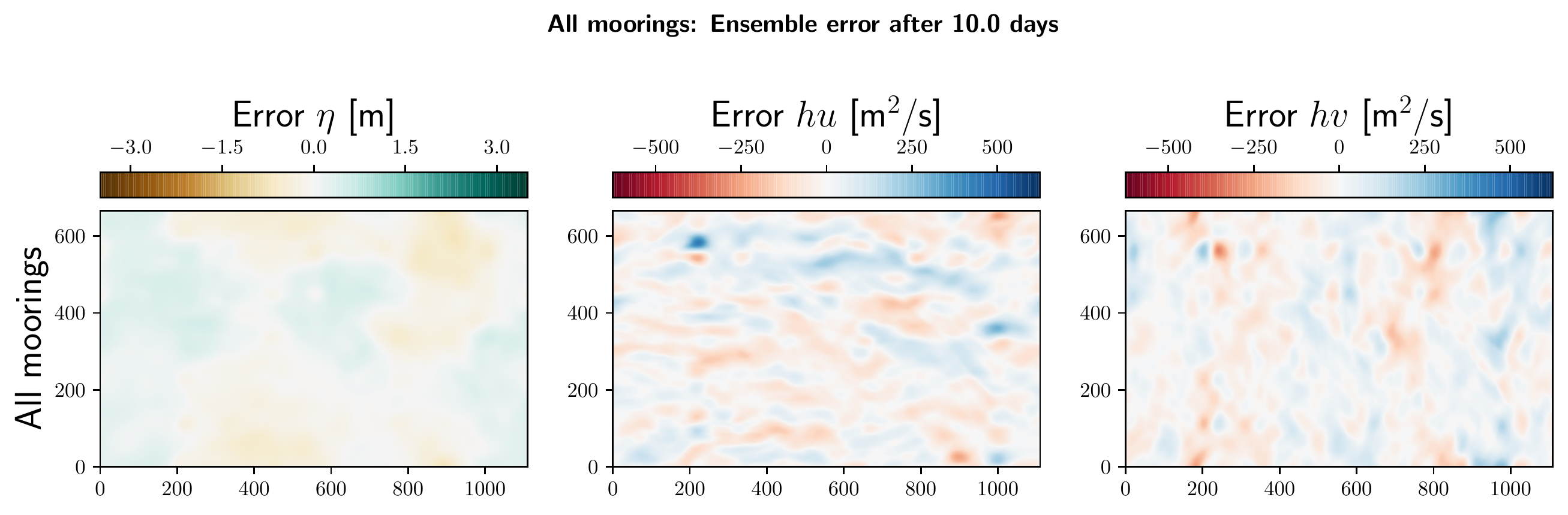}\\
        \rotatebox{90}{\small {\parbox{2cm}{(ii) LETKF \\ ($\phi=1.0$)}}} \quad
        \includegraphics[width=0.8\textwidth, trim=0.9cm 0.6cm 0cm 3.5cm, clip]{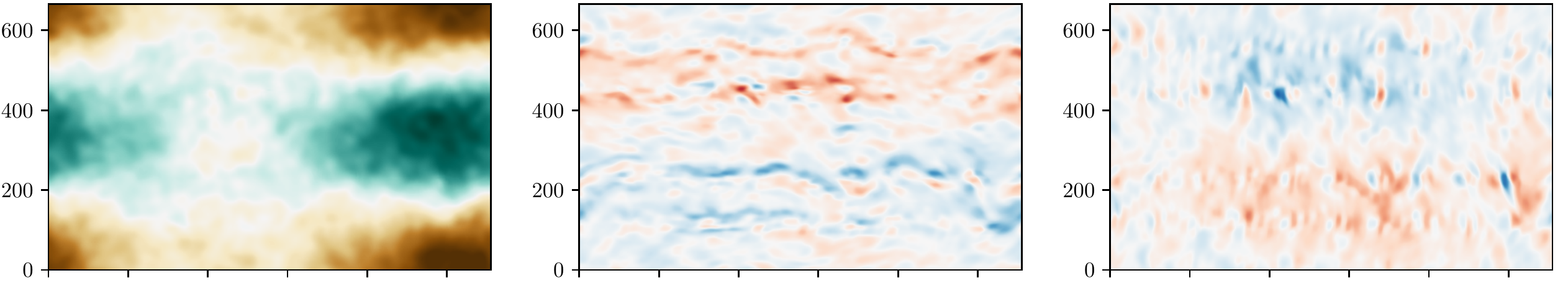}\\
        \rotatebox{90}{\small {\parbox{2cm}{(iii) LETKF \\ ($\phi=0.5$)}}} \quad
        \includegraphics[width=0.8\textwidth, trim=0.9cm 0cm 0cm 3.5cm, clip]{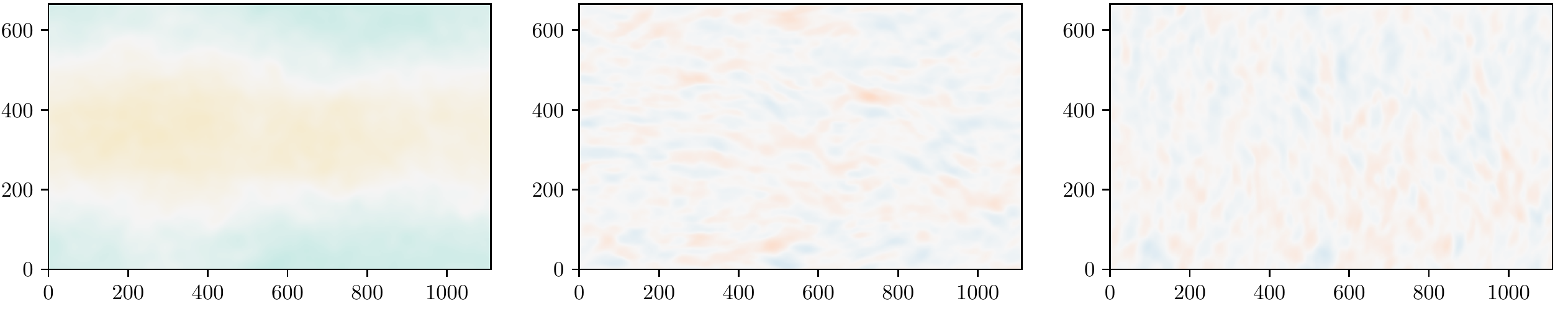}
        \caption{\quad Error $\textbf{err}_\text{mean}^{\text{day10, true}}$}
        \label{fig:gpuocean-error-mean}
    \end{subfigure}\vspace{1cm}
    \hfill
    \begin{subfigure}{0.8\textwidth}
        \centering
        \rotatebox{90}{\small {\parbox{2cm}{(i) IEWPF \\ \textcolor{white}{()}}} } \quad
        \includegraphics[width=0.8\textwidth, trim=0.9cm 0.6cm 0cm 1.5cm, clip, valign=b]{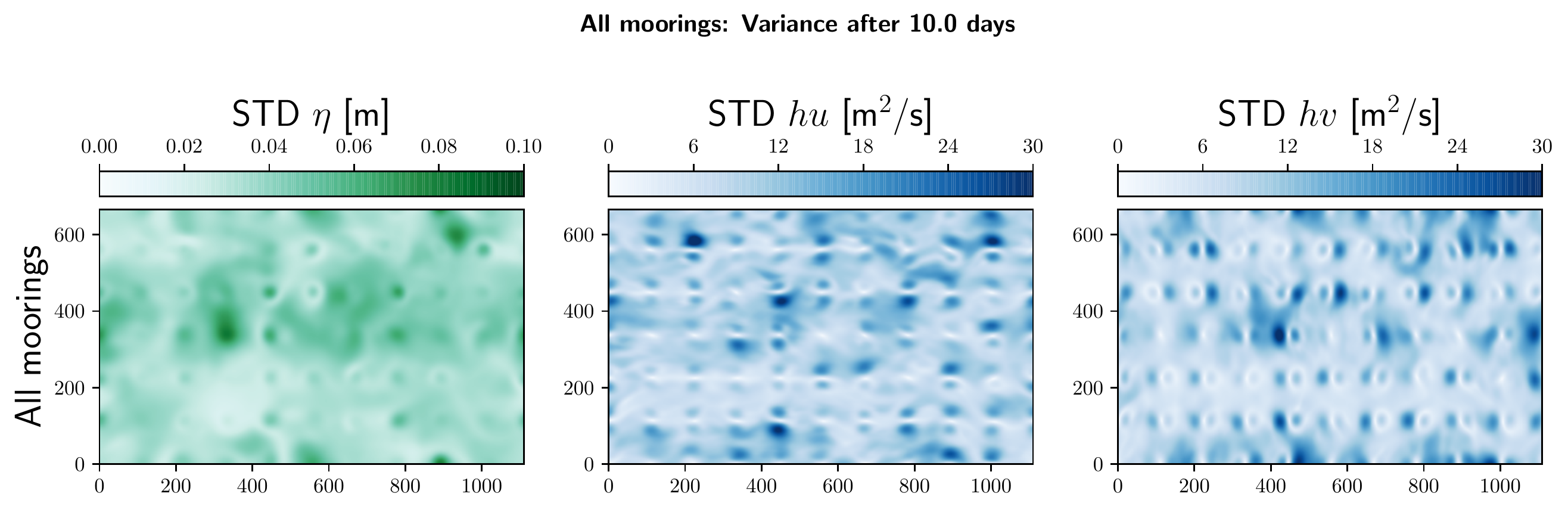}\\
        \rotatebox{90}{\small {\parbox{2cm}{(ii) LETKF \\ ($\phi=1.0$)}}} \quad
        \includegraphics[width=0.8\textwidth, trim=0.9cm 0.6cm 0cm 3.5cm, clip]{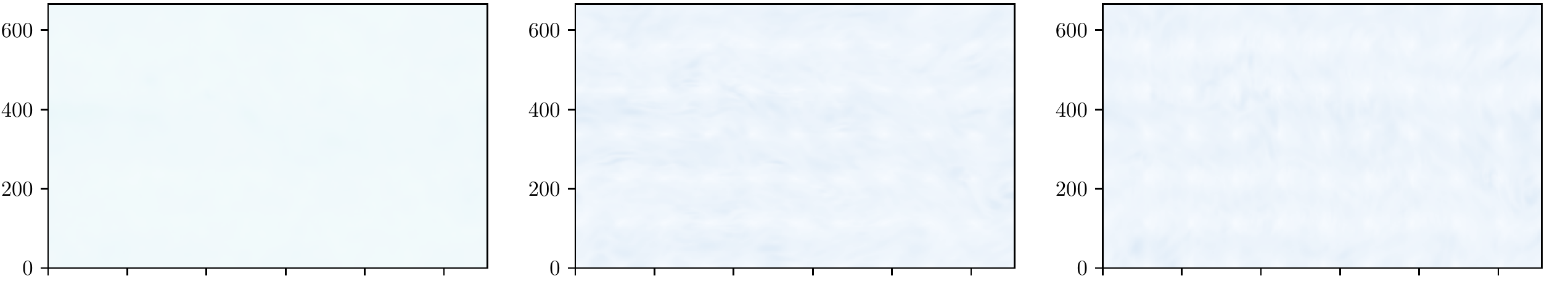}\\
        \rotatebox{90}{\small {\parbox{2cm}{(iii) LETKF \\ ($\phi=0.5$)}}} \quad
        \includegraphics[width=0.8\textwidth, trim=0.9cm 0cm 0cm 3.5cm, clip]{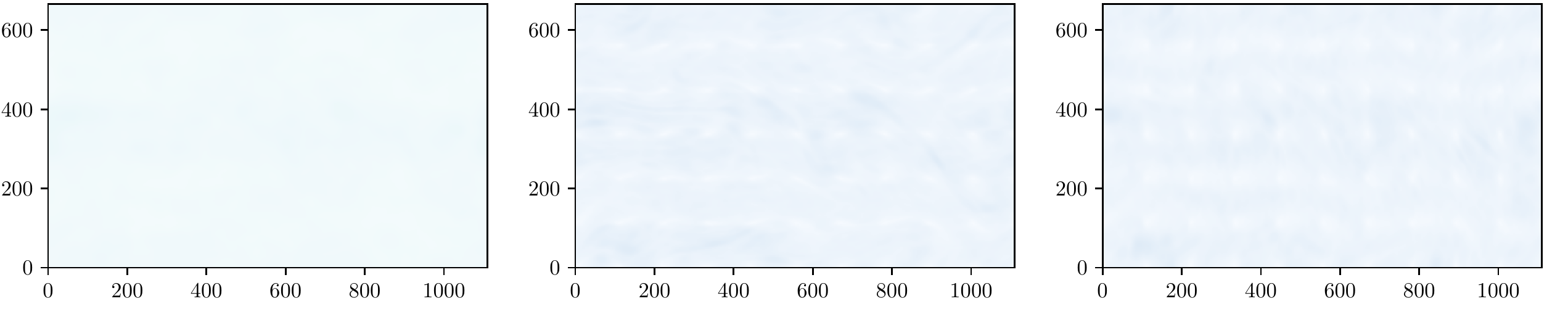}
        \caption{\quad Standard deviation $\text{STD}^\text{day10}$}
        \label{fig:gpuocean-error-std}
    \end{subfigure}
    \caption{Properties of the state estimation for the different physical variables in this simplified ocean model (sea-surface elevation $\eta$ as well as momenta $hu$ and $hv$) after day 10 measured in the error of the mean versus the truth and the standard deviation presented for the IEWPF as well as LETKF without and with inflation.}
    \label{fig:gpuocean-error}
\end{figure}

\Cref{fig:gpuocean-error-mean} shows the mismatch between the truth and the ensemble means of the conserved variables after assimilating the final observations on day 10. Significant differences become clear in the error of the sea-surface elevation $\eta$ (left): While the IEWPF has some moderate, relatively smooth error over the entire domain, the mean of the LETKF is far off in half of the domain.
In particular, the rims in the error field are very sharp, also recognisable in the error spots of the currents at the edges of the jets. This indicates that the ensemble produces very fast changing ocean fields with the tendency to non-physical members. 
However, inflation with $\phi=0.5$ (bottom) impressively fixes some of those issues and the error fields become much smoother and closely calibrated, even though there is still a recognisable, but weak inherited pattern in the error for elevation. 

In the standard deviations for the LETKF in \Cref{fig:gpuocean-error-std} (middle row), there is an expected pattern of low values around observation sites. Since the localisation only corrects around the buoys and leaves the forecast otherwise unchanged, the variance in one data assimilation step is mainly reduced in local areas. With the dynamical model over time, the variance reduction is disseminated over the entire domain. Furthermore, the standard deviation in the LETKF is on a very low level. 
Having areas of low error together with the sudden changes towards big errors suspects overfitting.
Taking a close look, the inflation aligns the differences of standard deviations between observed and unobserved areas a bit (lower row), but the variance is still kept on a low level.

There are structured artefacts identifiable around the observation sites for IEWPF in \Cref{fig:gpuocean-error-std}. Even though \Cref{fig:gpuocean-error-mean} shows that the ensemble mean gives a very precise description of the ground truth, the ensemble variance is large. As discussed in \citet{Holm2020c}, the IEWPF updates the momentum locally by inducing a corrective current formed by the structure in the model error covariance matrix $\modelErrorCov$. In this case, $\modelErrorCov$ induces geostrophically balanced dipole structures, which means that while improving the state at the observation site, we risk deteriorating the solution in its vicinity. This illustrates a weakness of IEWPF, showing that its quality is only as good as the structure of $\modelErrorCov$. By considering the extreme case with $\modelErrorCov = \obsErrorCov = \mathbf{I}$, \cref{eq:optStateUpdate} would reduce to only updating the variables that are observed while leaving all unobserved variables according to the forecast distribution. 

\paragraph{Drift trajectory forecasting}

To further compare the practical applicability of IEWPF and LETKF, we look at forecasts of drift trajectories starting at day 10.

\begin{figure}
    \centering
    \begin{subfigure}{0.8\textwidth}
        \centering
        \rotatebox{90}{\small {\parbox{2cm}{(i) IEWPF \\ \textcolor{white}{()}}} } \quad
        \includegraphics[width=0.8\textwidth, trim=0.9cm 0.6cm 0cm 2cm, clip]{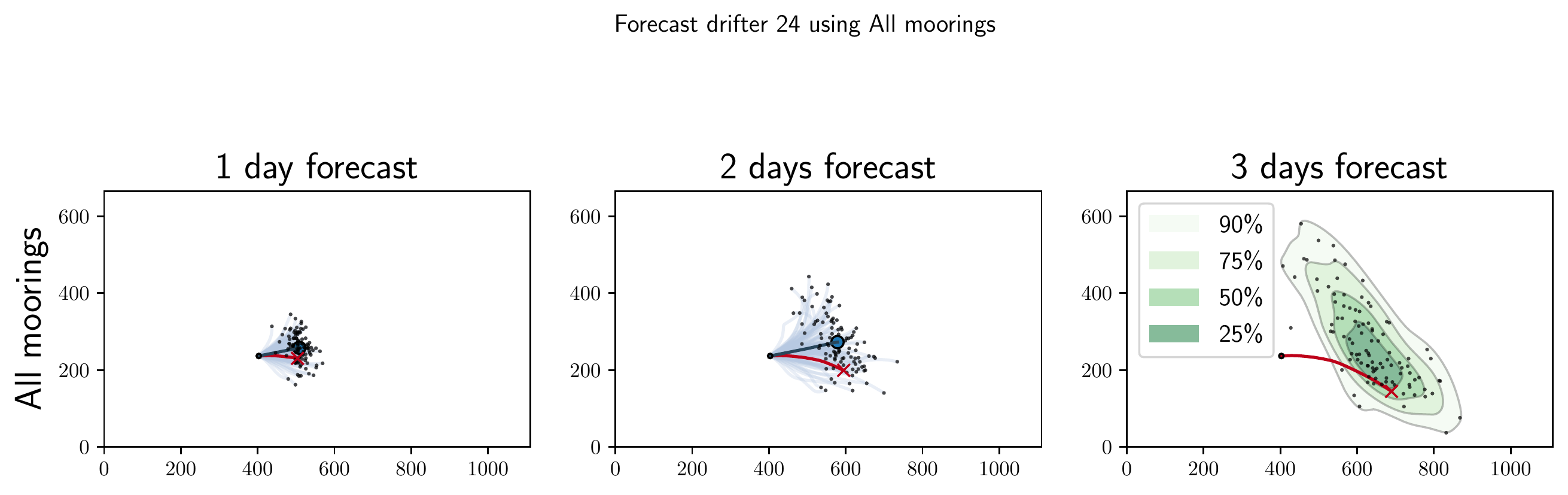}\\
        \rotatebox{90}{\small {\parbox{2cm}{(ii) LETKF \\ ($\phi=1.0$)}}} \quad
        \includegraphics[width=0.8\textwidth, trim=0.9cm 0.6cm 0cm 3.25cm, clip]{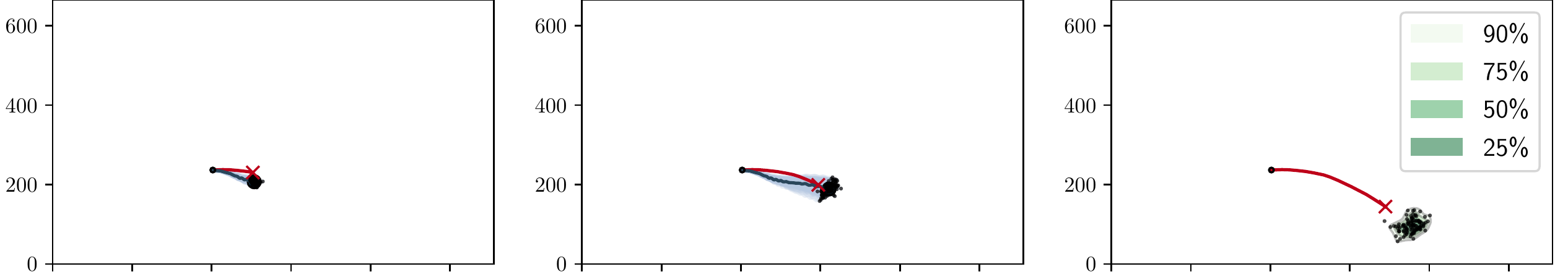}\\
        \rotatebox{90}{\small {\parbox{2cm}{(iii) LETKF \\ ($\phi=0.5$)}}} \quad
        \includegraphics[width=0.8\textwidth, trim=0.9cm 0cm 0cm 3.25cm, clip]{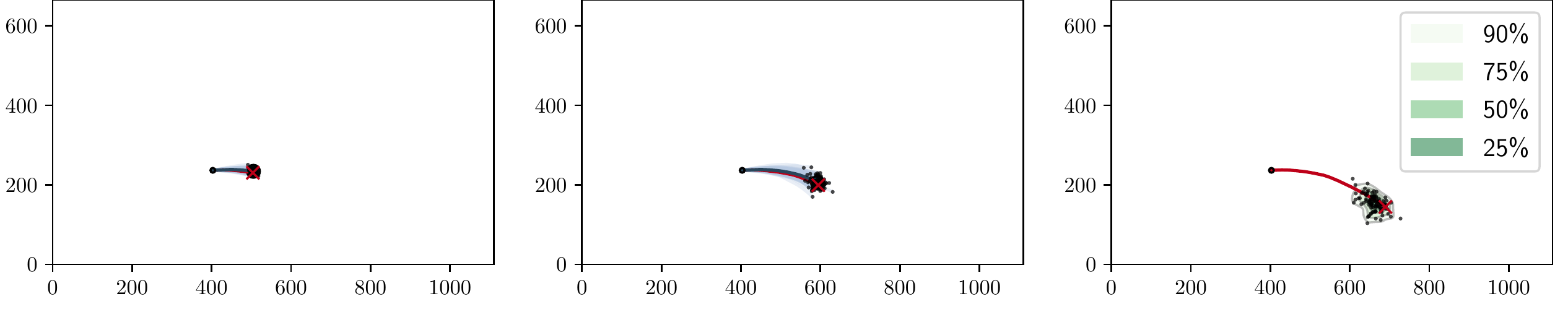}
        \caption{\quad Drifter 1}
    \end{subfigure}%
    \hfill 
    \begin{subfigure}{0.8\textwidth}
        \centering
        \rotatebox{90}{\small {\parbox{2cm}{(i) IEWPF \\ \textcolor{white}{()}}} } \quad
        \includegraphics[width=0.8\textwidth, trim=0.9cm 0.6cm 0cm 3.25cm, clip]{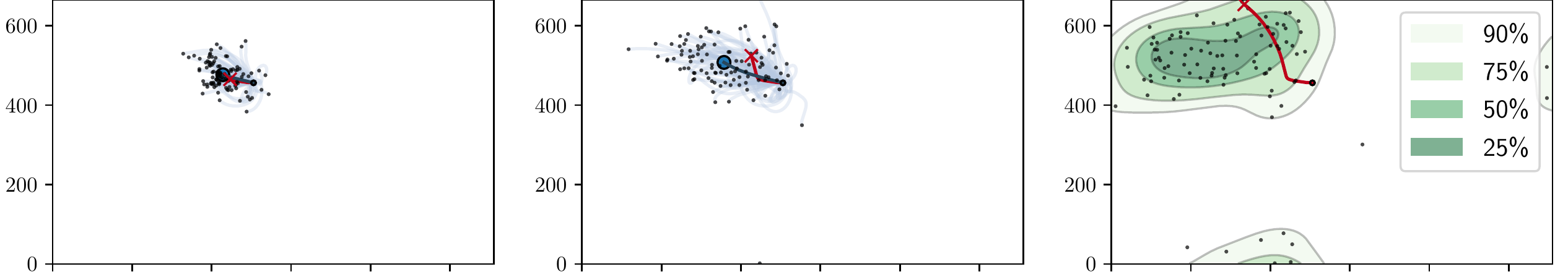}\\
        \rotatebox{90}{\small {\parbox{2cm}{(ii) LETKF \\ ($\phi=1.0$)}}} \quad
        \includegraphics[width=0.8\textwidth, trim=0.9cm 0.6cm 0cm 3.25cm, clip]{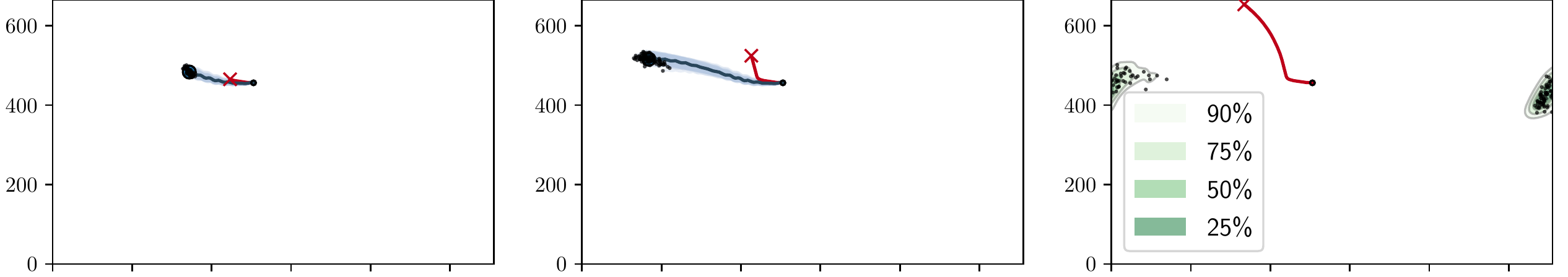}\\
        \rotatebox{90}{\small {\parbox{2cm}{(iii) LETKF \\ ($\phi=0.5$)}}} \quad
        \includegraphics[width=0.8\textwidth, trim=0.9cm 0cm 0cm 3.25cm, clip]{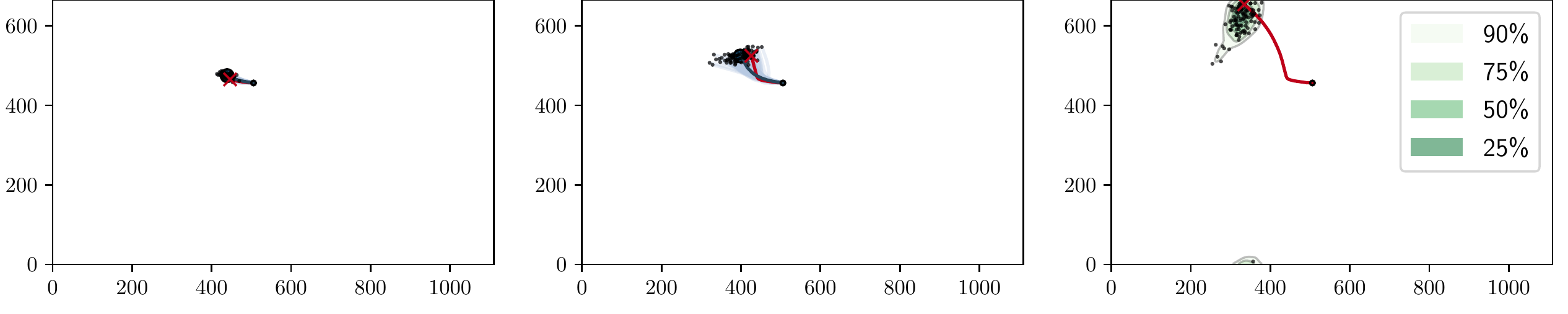}
        \caption{\quad Drifter 2}
    \end{subfigure}%
    \hfill 
    \begin{subfigure}{0.8\textwidth}
        \centering
        \rotatebox{90}{\small {\parbox{2cm}{(i) IEWPF \\ \textcolor{white}{()}}} } \quad
        \includegraphics[width=0.8\textwidth, trim=0.9cm 0.6cm 0cm 3.25cm, clip]{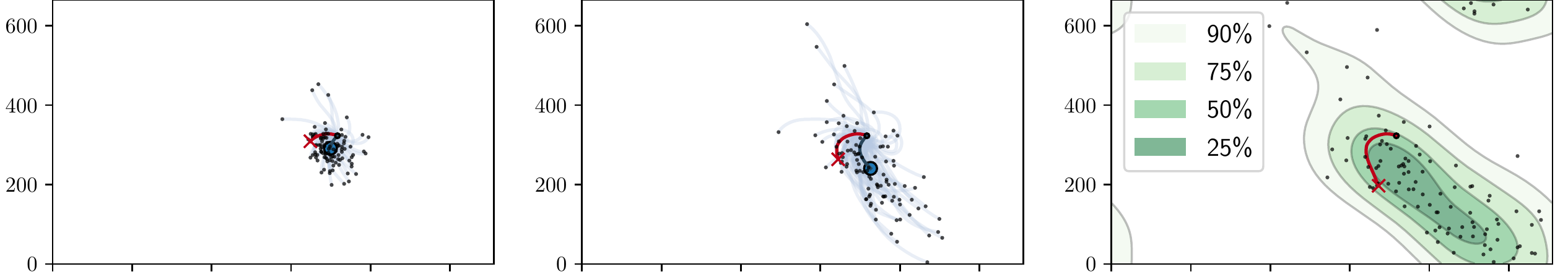}\\
        \rotatebox{90}{\small {\parbox{2cm}{(ii) LETKF \\ ($\phi=1.0$)}}} \quad
        \includegraphics[width=0.8\textwidth, trim=0.9cm 0.6cm 0cm 3.25cm, clip]{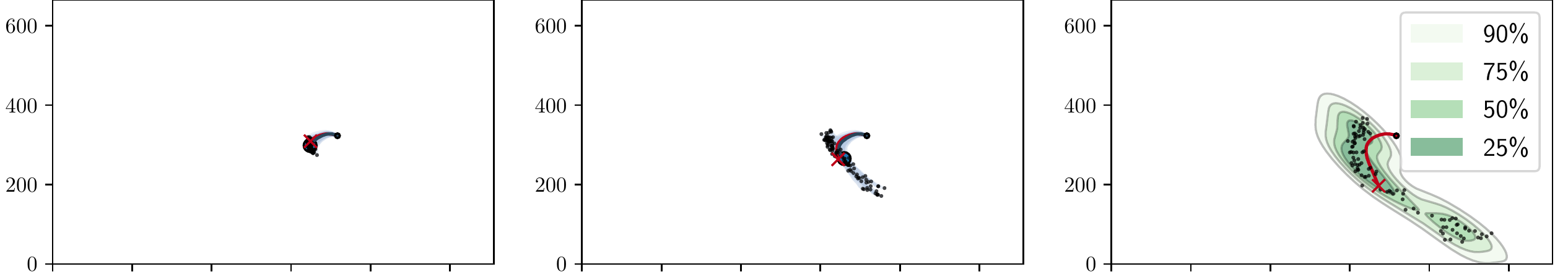}\\
        \rotatebox{90}{\small {\parbox{2cm}{(iii) LETKF \\ ($\phi=0.5$)}}} \quad
        \includegraphics[width=0.8\textwidth, trim=0.9cm 0cm 0cm 3.25cm, clip]{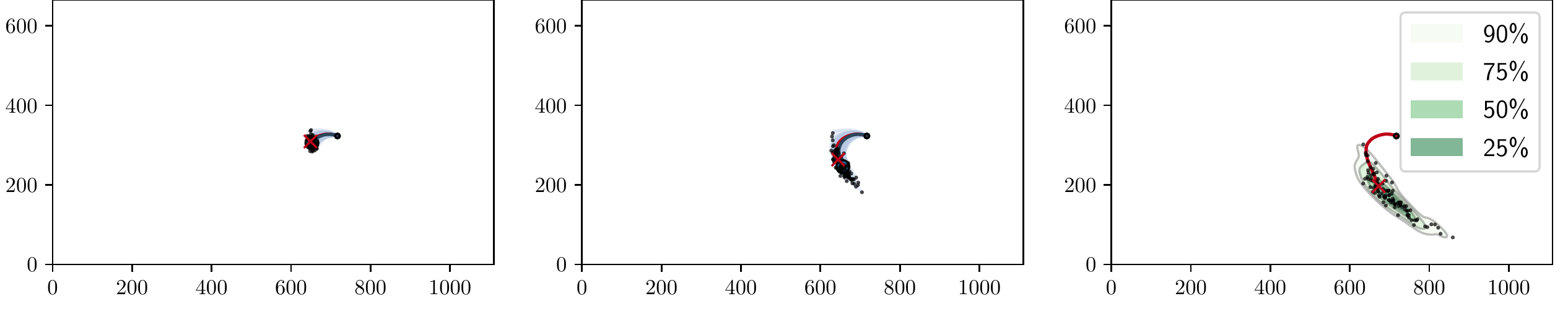}
        \caption{\quad Drifter 3}
    \end{subfigure}
    \caption{Drift trajectory forecasts for three different starting positions. True trajectories represented in red, and for the first two days trajectories of ensemble members are light blue and the ensemble mean in dark blue. For the third day, the forecasted drifter positions of all ensemble members are shown with black dots and selected levels of the estimated kernel density are visualised.}
    \label{fig:gpuocean-drift}
\end{figure}

\Cref{fig:gpuocean-drift} demonstrates the forecasted trajectories of drifters that are realised after ten simulation days in the simplified ocean model. 
The three drop locations are selected to capture different characteristics in the currents: Drifter 1 (display (a)) starts in the middle of a rather weak and big east stream. Drifter 2 (display (b)) starts in a rather strong west stream and drifter 3 (display (c)) starts in a turbulent area in between the dominating streams.
For the first two days of forecast, we show the true trajectory along with the trajectories for all ensemble members and the ensemble mean, whereas for the third day (right) we show the estimated kernel density \cite{Scott1992} of the final drifter locations along with the true trajectory.

For drifter 1, all trajectories have an east-wards drift, but the IEWPF members fan out from the beginning while the LETKF trajectories stay close together. Without inflation, the truth becomes an outlier in the LETKF forecast. With inflation, the truth stays within the forecast.
The trajectories from the IEWPF catch the truth in a high-probability area, but their spread covers almost the entire extent of the domain in the $y$-direction.

Even though drifter 2 starts within a jet, it drifts only shortly west-wards before it takes a sudden turn towards the north. Here, we can again see the turbulent behaviour of this non-linear model.
The trajectories of the IEWPF again spread out widely, and therefore does not reveal any consistent dynamical pattern in the underlying currents.
LETKF not only misses the true trajectory completely, it also shows some wriggling trajectories which indicates that there are unbalanced gravitational waves in the ensemble.
Inflation visibly increases the spread in the LETKF trajectories, and most of the ensemble members capture the sudden turn in the truth, even though this happens a day after assimilating the final observations.

Drifter 3, which is released in an unstable area, follows what is almost a rotation-like pattern.
Here, the IEWPF is unable to estimate a clear direction even for the first 24 hours, and after day 10 the drifter distribution stretches out across almost half the simulation domain.
In contrast, LETKF with and without inflating the ensemble gives a precise forecast for the first day, only showing a spread for the two last days. The truth is well represented by the ensemble for both experiments, but we see that the spread is remarkably reduced when using inflation.

In general, we see that even though IEWPF is able to give a good state estimation through the mean, the spread in the underlying ocean state is too large to facilitate precise drift trajectory forecasts. 
Furthermore, LETKF without inflation shows clear signs of overfitting, as the forecasts have low spread and do not match the ground truth.
Introducing inflation into the LETKF reduces this overfitting such that the true trajectories are correctly forecasted and uncertainty is better represented.
Even more important, giving more weight to the forecast that comes from the physical model prevents the ensemble from unintended anomalies. 
The drift trajectories estimation draws attention away from the ocean states towards dynamic visual characteristics in the ensemble.

\subsection{Discussion of skill scores}

Complementary to drift trajectory forecasts, we look into characteristics of both methods during the data assimilation phase between day 3 and day 10.
We compare statistical properties of the ensemble against the observation data. 
In this setting, the key idea of skill scores is to evaluate how reliably the ensemble can forecast the next observation.
An illustrative introduction with a lot of examples from atmospheric weather forecasting can be found in \citet[Chapter 7]{Wilks2005}.
Mathematically speaking, in this setting with data comparison, a score is
\begin{equation}
    s\left(\widehat{F}\DAstep[f],\obs\DAstep\right)\in\R,
\end{equation}
which in our case quantifies some property of the empirical distribution from an ensemble forecast $\left(\obsMatrix\state\DAstep[f]\member\right)_{e=1}^\ensembleSize$ against the true observation $\obs\DAstep$, meaning $hu_j$ and $hv_j$ for all $j=1,\dots,\obsSize$.
We consider three different skill scores to judge the performance.

\paragraph{Bias}
After asserting the calibration of the full analysis mean in \Cref{fig:gpuocean-error-mean}, we investigate this further by evaluating the bias of the forecast as 
\begin{equation}
    s_1\DAstep = \frac1\obsSize \sum_{j=1}^{\obsSize}\left[{\overline{hu}_j\DAstep[f] -\obs\DAstep_{j,1}  +  \overline{hv}_j\DAstep[f] - \obs\DAstep_{j,2}}\right].
\end{equation}
Here, $\overline{hu}$ and $\overline{hv}$ are the ensemble means. 
The bias discovers systematic trends off in the estimator. 

\paragraph{Mean square error}
We further investigate the distance of each ensemble member individually from the data by measuring the MSE as 
\begin{equation}
    s_2\DAstep = \frac1\ensembleSize \sum_{e=1}^{\ensembleSize}\left[{\frac1\obsSize \sum_{j=1}^{\obsSize}{|hu_{e,j}\DAstep[f] -\obs\DAstep_{j,1}|^2+ |hv_{e,j}\DAstep[f] - \obs\DAstep_{j,2}|^2}}\right].
\end{equation}
The MSE equals zero only when all ensemble members predict the observation exactly. 
However, this is of course not desired from a probabilistic forecast representing associated uncertainty.
Nevertheless, a small MSE is desired and yields accurate fit to the data respecting the standard deviation in the observation error.

\paragraph{Continuous ranked probability score}
Similar to the integrated quadratic differences which compared distribution forecasts, we use a scoring rule that analyses the distribution of the ensemble members with the observation, see \citet{Gneiting2007}.
The CRPS is here defined by
\begin{equation}
    s_3\DAstep = \frac1\obsSize \sum_{j=1}^\obsSize \left[{\frac1\ensembleSize \sum_{e=1}^{\ensembleSize}{|hu_{e,j}\DAstep[f] -\obs\DAstep_{j,1}|+ |hv_{e,j}\DAstep[f] - \obs\DAstep_{j,2}|} - \frac1{2\ensembleSize^2}\sum_{e=1}^{\ensembleSize}\sum_{k=1}^{\ensembleSize}{|hu_{e,j}\DAstep[f] - hu_{k,j}\DAstep[f]|+ |hv_{e,j}\DAstep[f] - hv_{k,j}\DAstep[f]|}}\right].
\end{equation}
Large CRPS values can originate from bias (first terms) or the spread in the ensemble (last terms).
Together with the scores for the bias and MSE, this allows one to identify the source of ensemble errors and to infer the properties of the ensemble.

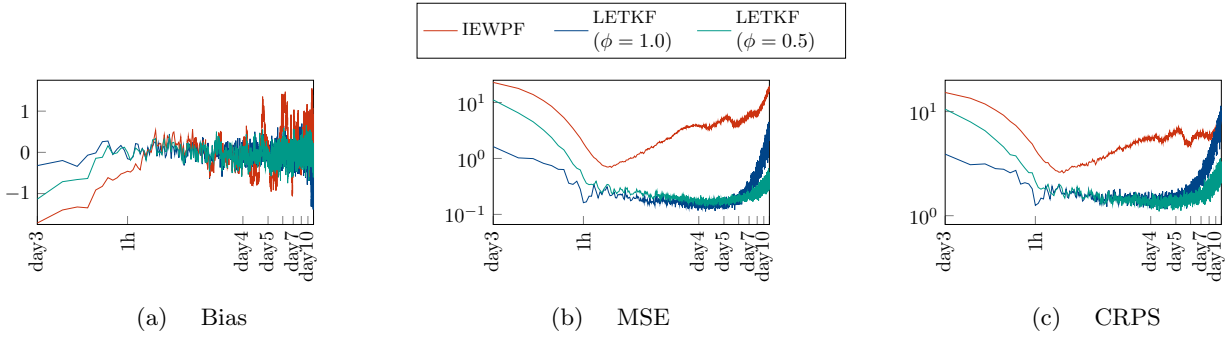
\begin{figure}
	\centering
	\begin{subfigure}[t]{0.3\textwidth}
		\begin{tikzpicture}[scale=0.75]
		\begin{axis}[
		width=0.95\textwidth,
		height=0.5\textwidth,
		scale only axis,
        legend pos=north west,
 		tick pos = left, 
 		xmin=1,
 		xmax=2016,
 		xtick={1,12,288,576,...,2016},
 		xticklabel style={rotate=90},
 		xticklabels={day3,1h,day4,day5,,day7,,day10},
        xmode=log,
 		ymin=-1.75,
 		ymax = 1.75,
 		yminorticks=false
		]
		
		\addplot[solid, color = pt2] table [x expr=\coordindex, y index=0,ignore chars={\#}] {files/scoresLETKF100.csv};
		
		\addplot[solid, color = pt3] table [x expr=\coordindex, y index=0,ignore chars={\#}] {files/scoresIEWPF.csv};
		
		\addplot[solid, color = pt4] table [x expr=\coordindex, y index=0,ignore chars={\#}] {files/scoresLETKF050.csv};
		
		\end{axis}
		\end{tikzpicture}
		\caption{\quad Bias}
	\end{subfigure}%
	\hfill
	\hspace*{-1cm}
	\begin{subfigure}[t]{0.3\textwidth}
		\begin{tikzpicture}[scale=0.75]
		\begin{axis}[
		width=0.95\textwidth,
		height=0.5\textwidth,
		scale only axis,
        legend pos=north west,
 		tick pos = left, 
 		xmin=1,
 		xmax=2016,
 		xtick={1, 12,288,576,...,2016},
 		xticklabel style={rotate=90},
 		xticklabels={day3, 1h,day4,day5,,day7,,day10},
        xmode=log,
        ymode=log,
 		ymin=0,
 		ymax = 25,
 		yminorticks=false,
 		legend style={at={(0.5,1.15)},anchor=south,cells={align=left}},
 		legend columns=3,
		]

		\addplot[solid, color = pt3] table [x expr=\coordindex, y index=1,ignore chars={\#}] {files/scoresIEWPF.csv};
		\addlegendentry{\small IEWPF~~~~}
		
		\addplot[solid, color = pt2] table [x expr=\coordindex, y index=1,ignore chars={\#}] {files/scoresLETKF100.csv};
		\addlegendentry{\small LETKF \\($\phi=1.0$)~~~~}

		\addplot[solid, color = pt4] table [x expr=\coordindex, y index=1,ignore chars={\#}] {files/scoresLETKF050.csv};
		\addlegendentry{\small LETKF \\($\phi=0.5$)~~~~}
		
		\end{axis}
		\end{tikzpicture}
		\caption{\quad MSE}
	\end{subfigure}%
	\hfill
	\begin{subfigure}[t]{0.3\textwidth}
		\begin{tikzpicture}[scale=0.75]
		\begin{axis}[
		width=0.95\textwidth,
		height=0.5\textwidth,
		scale only axis,
        legend pos=north west,
 		tick pos = left, 
 		xmin=1,
 		xmax=2016,
 		xtick={1,12,288,576,...,2016},
 		xticklabel style={rotate=90},
 		xticklabels={day3,1h,day4,day5,,day7,,day10},
        xmode=log,
        ymode=log,
 		ymin=0,
 		ymax =20,
 		yminorticks=false
		]
		
		\addplot[solid, color = pt3] table [x expr=\coordindex, y index=2,ignore chars={\#}] {files/scoresIEWPF.csv};
		
		\addplot[solid, color = pt2] table [x expr=\coordindex, y index=2,ignore chars={\#}] {files/scoresLETKF100.csv};
		
		\addplot[solid, color = pt4] table [x expr=\coordindex, y index=2,ignore chars={\#}] {files/scoresLETKF050.csv};
		
		\end{axis}
		\end{tikzpicture}
		\caption{\quad CRPS}
	\end{subfigure}%
	\caption{Evolution of skill scores for the IEWPF (red) as well as LETKF without inflation (blue) and with inflation (turquoise) for the data assimilation phase in the experiment.}
	\label{fig:gpuocean-skills}
\end{figure}

\Cref{fig:gpuocean-skills} presents the evolution of these skill scores for each data assimilation time step. These results are obtained from the same run as in \Cref{sec:gpuoceanNumResults}. Note that when we assimilate the first observation after spin-up on day 3, the spread in the ensemble is relatively large by construction for all methods. 
It should be noted that $hu$ and $hv$ take values up to \SI{500}{m^2/s}, which means that all methods have a relatively small bias.
In the starting phase, the LETKF immediately calibrates to the observations, whereas the IEWPF and inflated LETKF require several data assimilation steps to correct the bias. 
We see however, that the bias for all methods grow over time, but with the inflated LETKF keeping the smallest values.
The systematic bias from the inflation becomes negligible as it is sufficiently often reduced by repeated weighting with the unbiased analysis.

Looking at the MSE and CRPS, we see that both LETKF versions improve during the first few assimilation steps and stabilise at a certain level.
As expected, the initial improvement with inflation is slower than without inflation, but this gap is closed already after 1 simulation hour, which corresponds to 12 data assimilation cycles.
The quality of both LETKF versions are then stable until approximately simulation day 5, when the model dynamics gets more turbulent.
At this point, the solution without inflation starts to deteriorate due to the overfitting.
Note that we see a similar trend for the inflated LETKF later in the experiment.
By inflating even more ($\phi =0.25$, not shown), we confirm the trend with even slower convergence in the beginning and later divergence at the end.


Similarly to the inflated LETKF, IEWPF also converges during the initial data assimilation cycles, but the skill scores do not stabilise and instead diverge slowly.
The slow initial convergence was also pointed out in \Cref{sec:advectiondiffusion}, where we had to run the data assimilation sufficiently long to reach a stable level before being able to provide a fair comparison.

For a full assessment of the skills of a data assimilation method a single skill score gives only limited information. But for instance, the combinations of bias and CRPS broadens the insights, since the bias helps to explain the contributions in the CRPS. 
However, the differences especially between LETKF without inflation and IEWPF in the skill score results do not seem substantial, whereas we have seen contrary properties in the drift trajectories that stay concealed in the monitoring of the skill scores.
In general, this discussion tells us that the LETKF assimilates the ensemble much stronger towards data than the IEWPF and exemplifies the effects of inflation.

\paragraph{Rank histograms}
We next look at rank histograms to analyse the adequacy of the ensemble spread. A short time-span in the simulation is repeated multiple times and the rank of the simulation truth in the ensemble ordering is monitored at six dynamically independent locations.
Rank histograms then present the frequency of which a certain rank is reported among the $\ensembleSize$ realisations of the ensemble and the shape of the histograms is used as a diagnostic tool to identify shortcomings of methods \cite{Saetra2004}. 
Flat rank histograms are commonly understood as indication for ensemble consistency or reliability of the ensemble, as it means that every ensemble member is sampled from the same distribution as the truth.

\begin{figure}
    \centering
    \begin{subfigure}[htb]{0.3\textwidth}
        \includegraphics[width=0.8\textwidth]{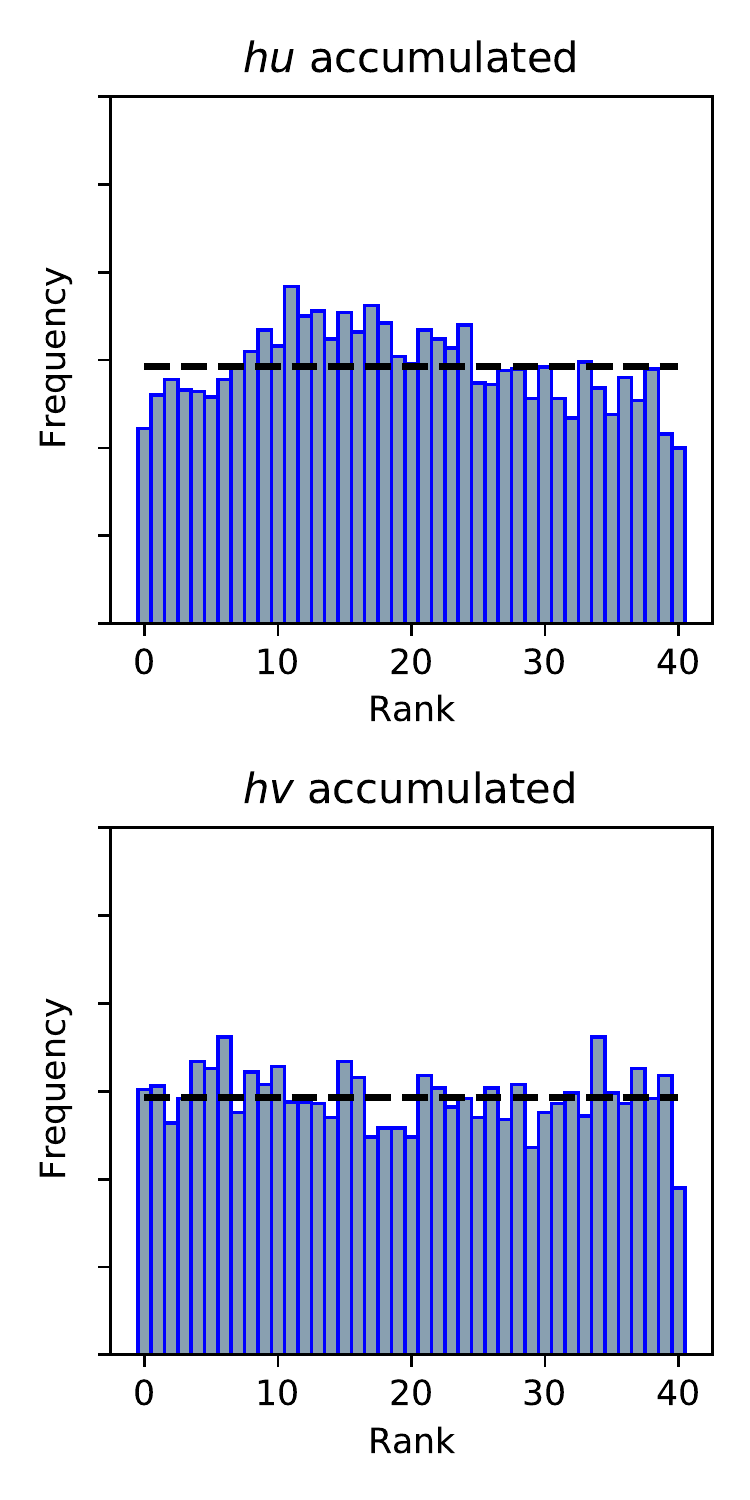}
        \caption{\quad IEWPF}
    \end{subfigure}
    \hfill
    \begin{subfigure}[htb]{0.3\textwidth}
        \includegraphics[width=0.8\textwidth]{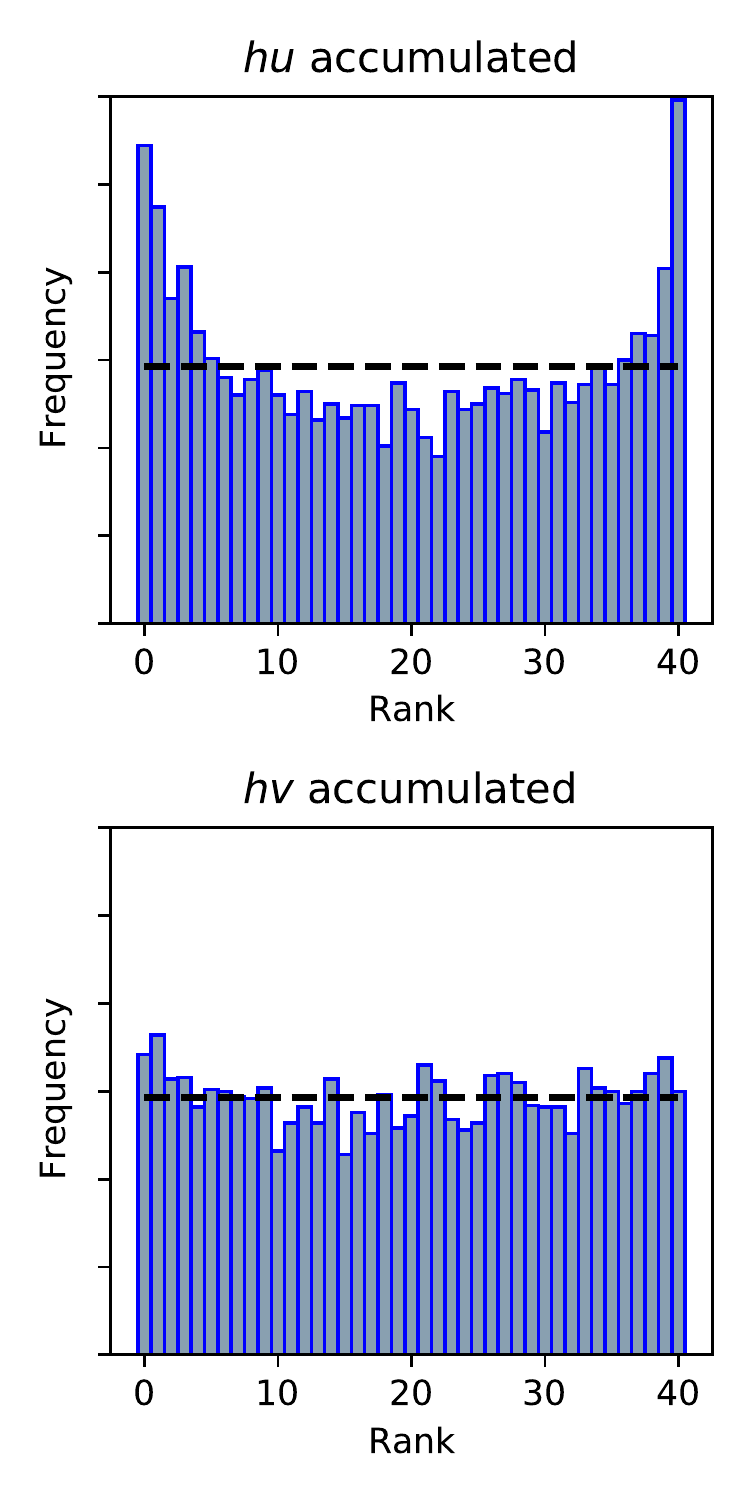}
        \caption{\quad LETKF ($\phi=1.0$)}
    \end{subfigure}
    \hfill
    \begin{subfigure}[htb]{0.3\textwidth}
        \includegraphics[width=0.8\textwidth]{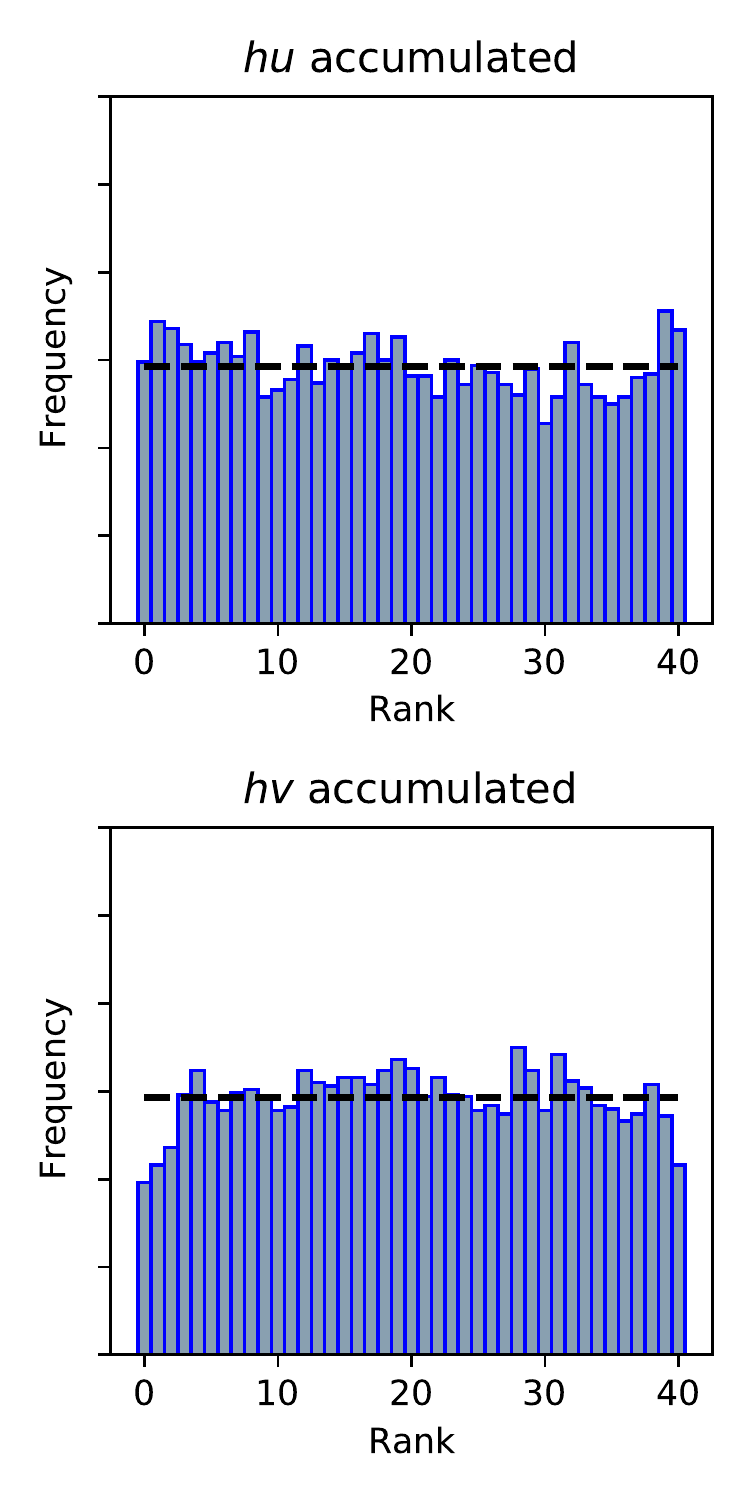}
        \caption{\quad LETKF ($\phi=0.5$)}
    \end{subfigure}
    \caption{Rank histograms recording the rank of the true observation within the ensemble for the observed variables. The dashed line indicates the hypothetical uniform distribution.}
    \label{fig:gpuocean-rankhist}
\end{figure}

In \Cref{fig:gpuocean-rankhist}, we show rank histograms from repeating our experiment 1000 times, using $\ensembleSize=40$ and simulating the first hour of data assimilation after the spin-up only.

The most striking result is the clear U-shape in $hu$ for LETKF without inflation, which indicates that the truth often is an outlier in the ensemble and that the ensemble is underdispersive. Furthermore, we observe that IEWPF produces a slight hill-shaped rank histogram for $hu$,  corresponding to an overdispersive ensemble.
Both these observations match well with what we saw in \Cref{sec:gpuoceanNumResults}.
In comparison, the $hu$ rank histogram for the inflated LETKF closely resembles a uniform distribution.
Note also that the rank histograms for $hv$ are flatter for all three methods, but with a slight tendency towards overdispersion for the inflated LETKF. 
This might be from the nature of the problem, as almost all dynamics in the case is along the $x$-axis.

While the rank histograms give insights how able the ensemble is to respect the uncertainty and we are able draw similar assertions from them as we suspected already before, \citet{Hamill2001} and \citet{Wilks2011} advice to be careful with their interpretation, since, e.g., spatial effects between the different locations become hidden.

\paragraph{Summary}
Based on these results for the nonlinear model, we see that the CRPS together with the bias are a good start for an analysis of the ensembles during the data assimilation phase, but by also analysing statistics over all state variables we are able to identify additional spatial artefacts and a fundamentally different variance in the ensemble.
Even though the rank histogram for the IEWPF looks reasonably flat and we get a well-calibrated mean for the state estimation, we see through the standard deviation that there are artifacts in the standard deviation, leading to a higher spread than what we see for LETKF.
This also made us realise how sensitive IEWPF is to the covariance structure in the model error.


Stress-testing the novel localisation scheme for LETKF in this high-dimensional non-linear experiments with very sparse data discloses that the LETKF has a tendency of overfitting to the observations, resulting in an underestimated variance in the ensemble.
LETKF is also not able to correctly estimate the unobserved variable $\eta$.
In practice, we see that inflating the ensemble through scaling the new local kernels is a good remedy for these flaws. It significantly weakens the defects, but still inherits the structures from the full LETKF.
We experienced that it is not necessary to fine-tune the inflation parameter as results were similar for $\phi = 0.25$ and $\phi=0.75$. 
We further point out that we tested classical covariance inflation, but this led to nonphysical states for $\eta$, while the variance of $hu$ and $hv$ was barely effected.
Hence, it is fair to use the LETKF with the proposed localisation strategy and the in-built inflation with caution.


\section{Conclusion}
\label{sec:conclusion}

We presented a novel localisation scheme for the LETKF applicable to spatially sparse point observations and we studied how its performance compares to the IEWPF, a state-of-the-art particle filter. We have considered two distinct cases, both motivated by simplified models applicable in oceanography. The first case studied state estimation of a linear Gaussian advection diffusion model. Here, the analytical filtering distribution was available for an in-depth statistical verification of the two methods in terms of estimation of the mean, covariances, distribution coverage, and spatial-temporal connectivity. In the assessment, which also included the standard ETKF, we recorded the performance of the ensemble-based methods in relation to the number of ensemble members and observation size. The second case was a non-linear shallow water model used for forecasting of drift trajectories. Here, we compared the performance of LETKF and IEWPF in terms of skill scores and forecast abilities. We also discussed inflation for the LETKF localisation scheme for this case. 
The extensive collections of comparison metrics allowed us to analyse plenty properties in ensemble representations.

Our results for the first case verified that both the IEWPF and LETKF with the proposed localisation scheme give very good estimates of the analytical reference solution. 
For moderate ensemble sizes, both methods delivered on par with the KF and clearly outperformed the ETKF in terms of RMSE and coverage probabilities. ETKF was best at estimating the covariance matrices, but it suffers from spurious correlations in the updates. 
The LETKF yields small divergences independent of the ensemble size. 
In the estimation of spatio-temporal model correlations, our results revealed that all three methods performed quite evenly. ETKF converges fast when the ensemble size grows.
Interestingly, we found that LETKF and IEWPF only showed minor improvements when increasing the ensemble size. IEWPF was the scheme benefiting the most from increased number of observations.

In the second non-linear case, we learned that both IEWPF and LETKF gave decent estimation of the momentum, but LETKF did so at the expense of the sea-surface level causing the drift trajectory forecast for some drifters to miss the truth altogether. These issues were also seen in the skill scores. IEWPF, on the other hand, showed minor artifacts around observation sites, indicating that the model error correlation matrix might not always represent the optimal mapping for assimilating the observations. 
In the case of LETKF, we showed that applying inflation to the weight kernel in the localisation scheme clearly reduced the sea-surface level calibration issues,
resulting in very good general performance. These results were backed up with high-quality results in the skill scores throughout the data assimilation period and precise predictions of the drift trajectories. 

To summarise, the most important findings in this paper can be listed as follows:
\begin{itemize}
    \item Our proposed localisation scheme for LETKF provides a computationally efficient method to assimilate very sparse point observations with good results even for relatively small ensemble sizes.
    \item A broader range of statistical evaluation metrics and skill scores reveals a deep insight into the assimilation methods. 
    \item We strengthen the argument that IEWPF, in contrast to most other PFs, is applicable to high-dimensional applications, but that the result is dependent on the structure of model error covariance matrix.
\end{itemize}


These results moreover opened up new directions for future research. For instance, it would be interesting to investigate how sensitive IEWPF is to the structure of the model error covariance matrix.
In principle, the suggested localisation scheme can be used with any version of an EnKF, and one may investigate whether it works equally well for other choices than the ETKF.
Beyond this, the sensitivity of the localisation concept to the inflation parametrisation could be tuned.
Furthermore, it would be interesting to test the version of LETKF in a real-world setting where the simulation model is only a simplification of the true model.

\appendix

\section*{Code Availability}

The source code used to produce the results presented in this paper is available under a GNU free and open source license in order to enhance scientific exchange \cite{Leveque2012}. 
For the time being, the code for \Cref{sec:advectiondiffusion} can be found under \url{https://github.com/FlorianBeiser/advectionDiffusion} and for \Cref{sec:gpuocean} under \url{https://github.com/florianBeiser/gpu-ocean/tree/enkf_doublejet}.

\section*{Acknowledgement}

The authors would like to thank Kjetil Olsen Lye for valuable feedback on the manuscript.




\bibliographystyle{apalike}

\end{document}